\documentclass[12pt]{article}

\usepackage{amsmath}
\usepackage{latexsym}
\usepackage{amssymb}

\newtheorem{thm}{Theorem}

\newtheorem{Defn}[thm]{Definition}
\newtheorem{Remark}[thm]{Remark}
\newtheorem{Note}[thm]{Note}

\newtheorem{Example}[thm]{Example}
\newtheorem{Examples}[thm]{Examples}
\newtheorem{Problems}[thm]{Problems}

\newtheorem{Problem}[thm]{Problem}
\newtheorem{Notation}[thm]{Notation}
\newtheorem{Number}[thm]{\!\!}

                  {\nopagebreak\hspace*{\fill}$\Box$\medskip\medskip\par}

\newcommand{\n}{\rm}

\newcommand{\N}{{\mathbb N}}
\newcommand{\R}{{\mathbb R}}

\newcommand{\Q}{{\mathbb Q}}

\newcommand{\C}{{\mathbb C}}
\newcommand{\K}{{\mathbb K}}

\newcommand{\g}{{\mathfrak g}}

\newcommand{\f}{{\mathfrak f}}
\newcommand{\oo}{{\mathfrak o}}
\newcommand{\gL}{{\mathfrak{gl}}}
\newcommand{\sP}{{\mathfrak{sp}}}
\newcommand{\e}{{\mathfrak e}}

\newcommand{\Gl}{\mathop{\rm Gl}\nolimits}

\newcommand{\Hom}{\mbox{\n Hom}}
\newcommand{\Aut}{\mbox{\n Aut}}
\newcommand{\End}{\mbox{\n End}}

\newcommand{\Sym}{{\mbox{\n Sym}}}

\newcommand{\SO}{\mbox{\rm SO}}
\newcommand{\UU}{\mbox{\rm U}}

\newcommand{\id}{\mbox{\n id}}

\newcommand{\m}{{\mathfrak m}}

\newcommand{\XX}{{\cal X}}
\newcommand{\ZZ}{{\cal Z}}
\newcommand{\YY}{{\cal Y}}
\newcommand{\II}{{\cal I}}

\newcommand{\MM}{{\cal M}}
\newcommand{\Gras}{{\rm Gras}}

\newcommand{\Flag}{{\rm Flag}}
\newcommand{\Fun}{{\rm Fun}}
\newcommand{\Herm}{{\rm Herm}}

\newcommand{\PP}{\Bbb{P}}
\newcommand{\HH}{{\cal{H}}}
\newcommand{\SSSS}{{\cal{S}}}

\newcommand{\PPP}{{\Pi}}
\newcommand{\SSS}{{\Sigma}}
\newcommand{\eps}{{\varepsilon}}
\newcommand{\OO}{{\rm O}}
\newcommand{\ad}{{{\rm ad}}}

\newcommand{\msk}{\medskip}
\newcommand{\ssk}{\smallskip}
\newcommand{\nin}{\noindent}

\begin{document}

\title{Is there a Jordan geometry underlying quantum physics?}

\author{Wolfgang Bertram\footnote{\footnotesize
Institut \'Elie Cartan Nancy, Nancy-Universit\'e, CNRS, INRIA, 
Boulevard des Aiguillettes, B.P. 239, F-54506 Vand\oe{}uvre-l\`{e}s-Nancy,
 France;
{\tt bertram@iecn.u-nancy.fr}  } }

\maketitle

\footnotesize{
{\noindent{\bf Abstract.\/}}
There have been several propositions for a geometric and essentially
non-linear formulation of quantum mechanics, see, e.g., 
 \cite{AS98},  \cite{BH01}, \cite{CGM03}, \cite{Ki79}. 
From a purely mathematical
point of view, the point of view of {\it Jordan algebra theory}
might  give new strength to such approaches:
 there is a ``Jordan geometry''
belonging to the Jordan part of the algebra of observables, in the same
way as Lie groups belong to the Lie part. Both the Lie geometry and the
Jordan geometry are well-adapted to describe certain features of quantum
theory. We concentrate here on the mathematical description of the Jordan
geometry and raise some questions concerning possible relations
with foundational issues of quantum theory.
}

\bigskip

{\footnotesize \noindent
{\bf AMS subject classification:}
{primary:
17C37, 
17C90, 
81R99, 
secondary:
53Z05, 
81P05} 



\bigskip

\noindent
{\bf Key words:}
\footnotesize{
 generalized projective geometries, Jordan algebras (-triple systems, -pairs),
quantum theory, twistor theory}

\bigskip


\section*{Introduction}

Can quantum theory be based on the commutative and non-associative
``Jordan product'' 
$$
X \bullet Y = {1 \over 2}(XY+YX)
$$ 
alone, or do
we need the associative product $XY$ somewhere in the background?
In his foundational
 work (\cite{J32}, \cite{J33}), Pascual Jordan gives an affirmative
answer to the first question. From a more contemporary perspective,
E.\ Alfsen and F.\ E.\ Schultz write (\cite{AlSch01}, p.\ vii): 
``...it has been proposed to model quantum mechanics
on Jordan algebras rather than on associative algebras \cite{JNW34}.
This approach is corroborated by the fact that many physically relevant
properties of observables are adequately described by Jordan constructs.
However, it is an important feature of quantum mechanics that the
physical variables play a dual role, as observables {\em and} as 
generators of transformation groups. ... Therefore both the Jordan
product and the Lie product of a $C^*$-algebra are needed for physics,
and the decomposition of the associative product into its Jordan part and
its Lie part separates two aspects of a physical variable.''

\ssk
I think that this point of view is very interesting and deserves
to be developed further. From the mathematical side, the ``Lie part''
has so far attracted much more attention than the ``Jordan part'', because
it has a beautiful relation with {\em geometry}, namely via
the {\em Lie functor}: to every (finite or infinite
dimensional)  Lie group we can assign
 a Lie algebra, which is a sort of infinitesimal neighborhood of
the origin of the group. 
Is there something similar for the ``Jordan part'' -- can we find a
global and geometric structure (finite or infinite dimensional) of which
the Jordan algebra somehow is an infinitesimal or tangent structure?
If this is so, one should expect that this structure might play a r\^ole
in physical theories and contribute to the understanding of 
``the geometry of quantum mechanics'' (I use this  term here
to embrace very different approaches such as, e.g.,  \cite{AS98} and \cite{Va85}).
In fact, this was the main motivation for the author's mathematical work on Jordan
structures, leading to the result that
there is indeed  a corresponding geometric object, 
 introduced in \cite{Be02} and called {\em generalized projective geometry}.
The first aim of the present work is to explain this purely mathematical
theory to readers coming from physics rather than from mathematics. 

\ssk The second aim is to raise the question whether there ought to be
relevant consequences of such a ``Jordan geometric''
approach  for physics. However, since the
author is a mathematician and not a physicist, 
I will only try to motivate why I think that this question may be
interesting, but not to answer it:
if the algebra of observables is indeed equivalent to some geometric,
``global'' and non-linear object, then it is possible to translate
the whole formulation from the linear level into a geometric and
non-linear language. As long as one restricts oneself to a faithful
translation, nothing is gained, and also nothing is lost.
Now,  all general arguments in favour
of geometric approaches, given,
 e.g., in \cite{AS98} and \cite{CGM03},  remain fully valid, and
as explained by these authors, the geometric formulation 
inevitably suggests new ideas and concepts which can no longer be
considered as a faithful translation of the theory we started with. 
In other words, at this point speculations begin.
If one believes that the present formulation of quantum theory is
complete, then of course one has no
reason and no need for such speculations.
For the sake of clarity I should admit that this is not my conviction, 
and I rather  adhere the point of view of several
authors,  explained very convincingly by R.\ Penrose in \cite{Pen05}
Chapters 29 and 30, that the present formulation is not satisfactory
and that there are foundational problems which 
are ``not just matter of philosophical interest'' (loc.\ cit., p.\ 865).
My speculations are, to some extent, similar to those of the
authors mentioned above, but in some parts they are different
and, perhaps,  complementary.
More specifically, in the first chapter of this work, I describe the general
features of Jordan geometries in an informal way, using terms borrowed
from the language of physics and thus suggesting a hypothetical physical
interpretation. The main features are:

\msk \nin
(1) {\em Duality.} The mathematically important distinction between space
and dual space, which also is a fundamental feature of Jordan theory
(cf.\ the notion of a {\em Jordan pair}, explained in Chapter 2), 
should also appear in a geometric
formulation of quantum theory; one may call it a duality between
``bras'' and ``kets'', or between ``observables'' and ``observers''.

\ssk \nin (2) {\em Linearity.} 
It has been strongly emphasized that quantum mechanics is a
{\em linear} theory -- and that,
if we  sacrifice linearity, this should be 
done in a ``subtle but essential way'' (\cite{AS98}, introduction). This is achieved
 by assuming a suitable {\em local linear structure} of our
geometries.

\ssk \nin (3) {\em Laws.}
The various local linear structures are related among each other via
algebraic laws, involving both the given geometry and its dual geometry.
A {\it generalized projective geometry} is a pair $(\XX^+,\XX^-)$ of
geometries that are locally linear and obey certain fundamental laws. 

\ssk \nin (4) {\em Polarities and energy.}
In classical
geometry, {\it polarities} are used to identify a projective
geometry with its dual projective geometry. In a likewise way, 
suitable identifications
between ``observables'' ($\XX^+$) and ``observers'' ($\XX^-$) in a generalized
projective geometry are called {\em polarities}. 
Physically, such a polarity seems to represent {\em energy} or the
{\em Hamilton operator}. However, although
this interpretation matches well with
the Jordan part of the usual observable ``Hamilton operator'', it matches
less well  with its Lie part, which is
related to the time evolution in quantum mechanics.
In other words, from a purely mathematical context we are lead to
a problem  looking quite similar to  the problem of the
 coexistence of the two quantum processes,
 the unitary ``$\bf U$-evolution'' and the ``state reduction
$\bf R$'' 
(here I use the labels introduced by R.\ Penrose, \cite{Pen89}, \cite{Pen05}).
 
\ssk \nin (5) {\em States and non-locality.}
As in non-commutative geometry, and
unlike the geometric approaches \cite{AS98}, \cite{BH01}, \cite{CGM03}, 
we started with
observables and not with {\it (pure) states}. Nevertheless, we can
associate a {\em geometry of states} to our geometry of observer-observables.
Here, a {\em state} is essentially a  global object of the geometry, something
like a projective line in a projective space or a light ray in compactified
Minkowski space; hence
the feature of non-locality is built in these concepts, and we feel that the
analogy with Penrose's twistor theory (cf.\ \cite{Pen05}) is not just an
accident.

\ssk \nin (6) {\em Geometry of special relativity.}
From the point of view of Jordan theory, the geometry of special 
relativity and the geometry of quantum mechanics are brothers,
the only difference being in dimension -- the former is associated
to Minkowski space (which is nothing but the Jordan algebra of Hermitian
$2 \times 2$-matrices) and the latter to the Jordan algebra of Hermitian
operators in an infinite dimensional Hilbert space.
Therefore everything we have said so far applies as well to the geometry
of special relativity (the conformal compactification of Minkowski
space). 
In particular, pure states in the geometry of special relativity indeed
lead to Penrose's twistor space.

\ssk \nin (7) {\em Hermitian symmetric spaces.}
On the one hand,
the importance of the ``complex Hermitian'' structure of quantum
mechanics has been emphasized, e.g., in \cite{CGM03} and in \cite{Pen05}.
On the other hand,
there is a well-known relation between {\em Hermitian symmetric spaces
of non-compact type}, also known as {\em bounded symmetric domains},
and certain (``positive Hermitian'') Jordan structures, see \cite{Up85}.
Our setting generalizes this correspondence in several regards.
Therefore, although at a first glance it looks rather different,
it has close relations to previous work of several authors relating
such structures to physics, and especially to quantum mechanics
(cf.\ \cite{Fr05}, \cite{Up87} and the extensive bibliography given in
\cite{Io03}). 

\msk
Throughout the text,
I try to illustrate all these concepts by simple examples from
linear algebra, so that the reader will more easily grasp (or be able
to skip) the formal
mathematical definitions which are given in Chapter 3. 
In Chapter 2, we give a  brief introduction to basic notions of
Jordan theory. The main result (equivalence of categories between
Jordan theory and generalized projective, resp.\ polar geometries) 
is stated in Subsection
3.6. Finally, in Chapter 4, I come back to the issue of possible
relations between physics and Jordan geometry  --
the least one can say is that some of its features 
match certain requirements on possible new approaches to the foundations
of quantum physics that have been put forward.
Moreover, the similarity with the geometry of special relativity (item
(6) above) may suggest how to carry such ideas even further, following the
ideas that have lead from special to general relativity.


\section{The general geometric framework}

\msk \nin
{\bf 1.1 Duality.} --
Not only in mathematics, but also in physics
it is useful to distinguish between a vector space $V$ and
its its dual space $V^*$, even if finally one wishes to identify
them. In Jordan theory, exactly the same phenomenon occurs: it turns out
to be useful to look at so-called ``Jordan pairs'' $(V^+,V^-)$ instead
of a single Jordan algebra $V$, even if one often  is interested in 
identifying $V^+$ and $V^-$ with $V$ as sets
 (see Chapter 2 for the formal definitions).

\ssk
Therefore we define our geometric ``universe'' as a {\it pair geometry}
$(\XX^+,\XX^-)$; this means just that $\XX^+$ and $\XX^-$ are sets, which we
call the {\it space of observables} and the {\em space of observers}, respectively,
such that there exists a basic {\em transversality relation}, 
denoted by $\top$:
a pair $(x,\alpha) \in \XX^+ \times \XX^-$ is called {\em transversal}, and
 we then also say that
 ``$\alpha$ can observe $x$'', and we then
write $\alpha \top x$ or $x \top \alpha$,  such that

\ssk
\nin (a) every observer can observe at least one observable:
for all $\alpha \in \XX^-$, there exists $x \in \XX^+$ with
$x \top \alpha$;

\nin(b) every observable can be observed by at least one observer:
 for all $x \in \XX^+$, there exists $\alpha \in \XX^-$ with
$x \top \alpha$.

\ssk
\nin
Here and in the sequel, all 
 assumptions will be such that we can turn things over: 
the r\^ole of $\XX^+$ and $\XX^-$
is entirely symmetric -- the pair $(\XX^-,\XX^+)$ is 
a universe with the same rights as  $(\XX^+,\XX^-)$, called its {\em dual 
universe}. Writing
$$
\alpha^\top := \{ x \in \XX^+ | \alpha \top   x \}
$$ 
for the set
of observables that can be observed by an observer
$\alpha$ (the ``visible world of $\alpha$''), our assumption means
that $\XX^+$ is covered by such sets, and vice versa.
In contrast, the set
 $\XX^+ \setminus \alpha^\top$, called
the {\it horizon} or the {\it infinite set of $\alpha$},
 may or may not be empty. 

\ssk
{\bf Example.} {\em A familar example of a pair geometry
 is given by a projective space
 $\XX^+ = \PP(W)$ and its dual projective space of hyperplanes
$\XX^-$ (which may be identified with $\PP(W^*)$,
where $W$ is a vector space and $W^*$ its dual
space), with
$x \top \alpha$ meaning that $x$ does {\em not} belong to the 
hyperplane $\alpha$;
in other words,
 $\alpha^\top$ is the complement of the hyperplane $\alpha$, and its horizon is
 the usual ``hyperplane at infinity''.
In the same spirit, one may consider the {\em Grassmann geometry
of type $E$ and co-type $F$}, 
$$
(\XX^+,\XX^-) = (\Gras_E^F(W),\Gras_F^E(W)),
$$
where $W=E \oplus F$ is a fixed direct sum decomposition of the vector space
$W$,
and $\Gras_A^B(W)$ denotes the set of all subspaces $Y$
in $W$ that are isomorphic to $A$ and such that $Y$ has some complement
that is isomorphic to $B$.
A pair $(U,V) \in \XX^+ \times \XX^-$ is transversal if and only if $U$ and
$V$ are  complementary 
subspaces: $W=U \oplus V$.
}

\msk \nin {\bf 1.2 Linearity.} -- 
The next structural ingredient to be added to our universe $(\XX^+,
\XX^-;\top)$ is the {\it principle of linearity}: for all observers $\alpha$,
the visible world $\alpha^\top$ is a linear space.
More precisely, fixing an arbitrary  observable $o \in \alpha^\top$ as
``origin'' in $\alpha^\top$, we require that a structure of a linear (i.e.,
vector) space with origin $o$ be given on $\alpha^\top$.
By duality, the same shall hold for $o^\top$ with origin
$\alpha$; if such a  structure is given for all transversal pairs
$(o,\alpha)$,
we say that $(\XX^+,\XX^-,\top)$ is equipped with a 
structure of {\it linear pair geometry}.
It may happen that the underlying affine space structure on
$(\alpha^\top,o)$ does not depend on the choice of the origin $o$; 
if this is alwas the case, the geometry is called an {\it affine pair
geometry}. In other words, $\alpha^\top$ then canonically carries the
structure of an affine space.
In any case, we think of $\XX^+$ as ``modelled on the linear
space $\alpha^\top$'', just as usual projective space $\R \PP^n$ is
modelled on usual affine space $\R^n$, and it is not misleading 
to picture $\XX^+$ as a, finite or infinite dimensional, smooth
manifold, covered by ``linear chart domains'' $\alpha^\top$ which in turn
are indexed by $\alpha \in \XX^-$.

\ssk
{\bf Example.} {\em Consider the Grassmann-geometry $(\Gras_E^F(W),
\Gras_F^E(W))$: it is a well-known exercise in linear algebra that
the set of complements of a given subspace carries canonically the structure
of an affine space, modelled on the linear spaces of linear operators
 $\Hom(F,E)$, respectively
$\Hom(F,E)$.  
Thus $(\Gras_E^F(W),
\Gras_F^E(W))$ is an affine pair geometry.

\ssk
More generally, one may consider the geometry $(\Flag_\e^\f,\Flag_\f^\e)$
of all flags in $W$ of a given ``type'' and ``cotype'', equipped with a
natural transversality relation; this defines a linear pair geometry
which, however, is no longer affine in general (cf.\ \cite{BL06}).}

\msk \nin
{\bf 1.2.1 Time.}
-- 
When we speak about vector spaces, we must specify a base field $\K$.
We consider $\K$ as ``time'', although  by no means we want by this to fix the
choice to be $\K=\R$, the real base field 
 -- some may prefer a ``complex time'',
or a ``$p$-adic time'' or yet another model (see \cite{Pen05}, Chapter 16
 for some 
remarks on the ``base field of physics''). Personally, I prefer 
models in which ``infinitesimal times'' exist, like the ring
$\R[\eps] = \R \oplus \eps \R$ ($\eps^2=0$)
 of {\it dual numbers}, where $\eps$ may be related in some mysterious
way to the Planck time. Thus, instead of fields, we will admit
also (commutative unital) base {\it rings}, and the term ``linear space''
means just `` $\K$-module''. 
By a fortunate coincidence of terminology, the unit group $\K^\times$
of $\K$ can then be seen as the set of possible ``units of time measurement'';
the non-invertible elements of $\K$ may be 
considered as ``infinitesimal times'', which cannot be used as units
of time measurement. 
Whatever the structure of time be, and
in contrast to a trend set by Hilbert's ``Foundations of Geometry''
(cf.\ \cite{Bue95}), we accept the base ring $\K$ as
God-given, and we do not try to ``reconstruct'' it from incidence structures
or other data.

\msk \nin
{\bf 1.2.2 Laws.} --
We assume that the universe $(\XX^+,\XX^-)$ is a linear or, better, affine
pair geometry over $\K$, governed by  {\it laws}.
These laws give it a certain structure, somewhat similar to the one
of a projective
geometry $(\PP \HH, \PP \HH^*)$ of a Hilbert space $\HH$, but
more flexible and incorporating many other situations. 
In a sense, these laws describe the ``basic rules of communication''
between various observers $\alpha, \beta$ and their visible worlds
$\alpha^\top$, $\beta^\top$ which, after all, shall be interpreted 
by $\alpha$ and $\beta$ as their images of the {\it same} world --
at least, if they live sufficiently close to each other so that 
the common part $\alpha^\top \cap \beta^\top$ of their visible worlds 
is non-empty.
The formal statement of such laws, to  be given in Chapter 3,
is described by {\em identities} for the so-called {\em structure maps}:
if $x \top \alpha$, $y \top \alpha$, $z \top \alpha$ and $r \in \K$, then let
$r_{x,\alpha}(y):= r y$ denote the scalar multiple $r \cdot y$, 
and $y+_{x,\alpha} z:=y+z$ the sum of 
$y$ and $z$ in the $\K$-module
$\alpha^\top$ with zero vector $x$. In other words, we define maps of three
(resp.\ four) arguments by
\begin{eqnarray*}
\PPP_r :=\PPP_r^+:&
 (\XX^+ \times \XX^- \times \XX^+)^\top  \to \XX^+, & 
(x,\alpha,y)   \mapsto \PPP_r(x,\alpha,y) :=r_{x,\alpha}(y),
\cr
\SSS:=\SSS^+: &(\XX^+ \times \XX^- \times \XX^+ \times \XX^+)^\top  \to 
\XX^+,  &
(x,\alpha,y,z)   \mapsto \SSS(x,\alpha,y,z) := y +_{x,\alpha} z,
\end{eqnarray*}
where the domain of definition of $\PPP_r$ is the ``space of generic triples'',
$$
 (\XX^+ \times \XX^- \times \XX^+)^\top  = 
\{ (x,\alpha,y) \in \XX^+ \times \XX^- \times \XX^+ \vert \,
x \top \alpha, \, y \top \alpha \},
$$
and the domain of $\SSS$ is the similarly defined space of generic quadruples.
By duality, $\PPP_r^-$ and $\SSS^-$ are defined.
The structure maps encode all the information of a linear pair geometry:
by fixing the pair $(x,\alpha)$, the structure maps describe the linear
structure of $(\alpha^\top,x)$, resp.\ of $(x^\top,\alpha)$.
In this way, linear pair geometries can be regarded as algebraic objects
whose structure is defined by (one or several) ``multiplication maps'', 
just like groups, rings or modules, and just like these they form a
{\em category} (as usual, morphisms are maps that are compatible with
the structural data).
In Chapter 3
 the class of {\em generalized projective geometries} will be singled out
by requiring certain identities for these structure maps.

\ssk {\bf Example.} {\em Again in the example of the Grassmann geometry,
one can give an explicit formula  for the structure maps in terms of
 linear algebra:
identifying  elements of $\Gras_E^F(W)$ with images of injective maps
$f:E \to W$, modulo equivalence under the general linear group
$\Gl_\K(E)$ ($f \sim f'$ iff $\exists g \in \Gl_\K(E)$:
$f'=f \circ g$), and elements of $\XX^-=\Gras_F^E(W)$ with  kernels
of surjective maps $\phi:W \to E$, again modulo equivalence under
$\Gl_\K(E)$, the basic transversality relation
is: $[f] \top [\phi]$, if and
only if $\phi \circ f:E \to E$ is a bijection. Then the structure map 
$\PPP_r$ is given by the explicit formula
$$
 \PPP_r([f],[\phi],[h])
= \big[ (1-r) f \circ (\phi \circ f)^{-1} + r h \circ (\phi \circ h)^{-1}
\big] .
$$
(Proof, cf.\ \cite{Be04}: first of all, note that the expression on the
 right hand  side is independent of the chosen representatives;
then choose new representatives such that $\phi \circ f = \id_E =
\phi \circ h$, and observe that this gives the usual formula of
a barycenter in an affine space.)
As in ordinary projective geometry, the affine picture in the model space
$\Hom(E,F)$ is given by writing
$f:E \to W = F \oplus E$ as ``column vector'' and normalizing the
second component to be ${\bf 1}_E$, the identity map of $E$, and similarly
for the ``row vector'' $\phi:F \oplus E \to E$ (see \cite{Be04}).
To get a feeling for the kind of non-linear formulas that appear in such
contexts, the reader may rewrite the preceding formula for $\PPP_r$
 by replacing
$f,\phi$ and $h$ by such column-, resp.\ row vectors, and then 
renormalize the right-hand side, in order to get the formula for the
multiplication map in the affine picture. The special case $r=\frac{1}{2}$
(the ``midpoint map'') is particularly important from a Jordan-theoretic
point of view.
}

\msk \nin
{\bf 1.2.3 Base points.} --
Although this may seem pedantic, we insist in clearly distinguishing between linear
pair geometries and those {\em with base point}: a {\em base point} is
a transversal pair, often denoted by $(o^+,o^-)$ or $(o,o')$, that is
chosen to be fixed ``once and for all'' (or, at least, until to the end of
the present sentence). Whereas geometric concepts should be base point-free,
our description of the universe often uses them -- from our ant's
perspective  we often do not 
realize that our visible world is just a part of the whole universe,
and we take this part for the whole. This remark applies to special
relativity as well as to quantum mechanics.

\ssk {\bf Example.} {\em  We explain the last statement.
 This can be done both in the
context of abstract $C^*$-algebras or in the concrete realization 
on a Hilbert space $\HH$. For simplicity, let us start
with the latter (the more abstract setting will be considered in
Example 2 of 1.3.2): let
 $\HH$ be a finite or infinite dimensional complex Hilbert space.
Our geometry $(\XX^+,\XX^-)$ will be
a subgeometry of the Grassmann geometry $(\YY^+,\YY^-)=(\Gras_\HH^\HH(W),
\Gras_\HH^\HH(W))$ where $W=\HH \oplus \HH$. Note that here $\YY^+ = \YY^-$ as
sets. Now we define
 $\XX^+ = \XX^-$ to be the  {\em space of Lagrangian subspaces on $W$
for the Hermitian form } (of ``signature $(\infty,\infty)$'')
$$
\omega:W \times W \to \K, \quad
\omega((u,v),(u',v'))=\langle u,v' \rangle +
\langle v,u' \rangle.
$$
In other words, $(\XX^+,\XX^-)$
 is the subgeometry of $(\YY^+,\YY^-)$ fixed under
the involutive
 automorphism ``orthocomplementation w.r.t.\ $\omega$'' (by definition,
a Lagrangian subspace $E$ is such that $E=E^\perp$). As in the preceding
examples, it is an exercice in linear algebra to show that this affine
pair geometry is modelled on the space of Hermitian operators
$\Herm(\HH)$.

\ssk
The usual framework of quantum mechanics is simply obtained by fixing a
base point $(o,o')$ in the Lagrangian geometry $(\XX^+,\XX^-)$, where
$o'$ singles out an affine part $\Herm(\HH)$ in $\XX^+$ in which $o$
is the zero vector. Of course, $\HH$ should then be infinite dimensional.
If $\HH$ is finite dimensional, say $\HH=\C^n$, then $\XX^+$ and $\XX^-$
are homogeneous spaces under the  action of the projective pseudo-unitary group
$\PP\UU(n,n)$, with stabilizer $P$ of $o$ being a certain maximal parabolic
subgroup, so that $\XX^+ \cong \PP \UU(n,n)/P$ is modelled on the (Jordan algebra of)
Hermitian $n\times n$-matrices $\Herm(n,\C)$. Now, for $n=2$, this Jordan
algebra is isomorphic to
 Minkowsi space $\R^{(3,1)}$, the group $\PP \UU(2,2)$ is isomorphic
to the conformal group $\SO(4,2)$ of Minkowski space, and $\XX^+ \cong
\SO(4,2)/P$ is precisely the {\em conformal compactification of Minkowski space}.
Therefore our geometric setting of quantum mechanics can be seen as the
infinite dimensional analog of the geometric, conformal completion of 
usual, flat special relativity.  The fact that quantum mechanics and
special relativity appear as {\em linear} theories corresponds to the
fact that a base point $(o,o')$ has been fixed.
}

\msk \nin
{\bf 1.3 Energy.} --
As long as the two dual worlds $\XX^+$ and $\XX^-$ remain neatly separated,
we are in the realm of ``projective geometry'', which is a beautiful theory,
but lacks the dynamics that we are used to from the real world.
In order to create dynamics, 
we must introduce some sort of identification between
$\XX^+$ and $\XX^-$ which, mathematically, is modelized by a pair of bijections
$p^+:\XX^+ \to \XX^-$, $p^-:\XX^- \to \XX^+$.
Henceforth, the observable $x$ and the observer $p^+(x)$ will be considered
as ``the same thing'', and similarly the observer $\alpha$ and the observable
$p^-(\alpha)$ are the same thing. Consistency requires then that $p^-$ 
is the inverse transformation of $p^+$, and moreover that $(p^+,p^-)$
respects the laws mentioned above (i.e., it defines an isomorphism of the
geometry onto its dual geometry).
Geometrically, this corresponds to what is sometimes called
a {\it correlation} in projective geometry, but from the point of view of physics,
it seems that it really is some sort of {\it energy}. Since it
 acts as a transformation, one might be tempted to
 use also the term {\it Hamilton operator}
for $p^+$ or for its inverse $p^-$.

\msk \nin
{\bf 1.3.1 Polarities.} --
There is something special about  humain beings, namely  that {\em
we can observe ourselves}. Let us call an observer {\it active} or
{\it non-isotropic} if it has this property, i.e., if
$\alpha \top p^{-}(\alpha)$, and {\it passive} or {\it isotropic} else.
In order to create dynamics, we must require that at least some
active observers shall exist; then our correlation $(p^+,p^-)$ is called a
{\em polarity}. In the remainder, we will mainly be concerned
with the ``active universe'' 
$\MM^{(p)} = \{ x \in \XX^+ \vert \, x \top p^+(x) \}$.
We do not require that the whole universe is active -- this may happen for
very strong energies which we call {\it elliptic polarities}, but in
general, it seems that polarities of hyperbolic type are more interesting
since singular points, such as possible ``beginnings'' or ``ends of the
universe'', will have to be passive.

\ssk {\bf Example.} {\em Polarities of projective spaces are constantly used
in classical geometry:
assume $W$ is a real
 Hilbert space; then one identifies a line $[x] \in \XX^+ =\PP(W)$
with its orthogonal hyperplane $[x]^\perp \in \XX^- = \PP(W^*)$. Since a 
scalar product is positive, there are no isotropic vectors: the polarity
is elliptic. But we may also work with general non-degenerate forms
(symmetric or skew-symmetric) and then get more general polarities.  
If we work over complex Hilbert spaces, then the scalar product induces
a $\C$-antilinear polarity (so we are working with complex geometries, considered
as real ones); there are of course also $\C$-linear polarities,
coming from non-degenerate $\C$-bilinear forms on $W$, but their polarities
always have isotropic elements. It is clear that such kinds of 
``orthocomplementation polarities'' can be defined also for Grassmann
geometries (provided that the $\K$-module $W$ admits suitable non-degenerate
bi- or sesquilinear forms). In any case, the affine picture of such kinds
of polarites is given by identifying a linear operator from $\Hom(E,F)$
with a suitable adjoint in $\Hom(F,E)$.   }

\msk \nin
{\bf 1.3.2 Null-systems.} --
On the other hand, it is theoretically possible that energies are so weak
that they admit no active observables whatsoever; one would call them
{\it null-energies} or {\it null-systems}. 
It is even theoretically conceivable that there  be an
{\it absolute null-energy} which is defined to be a pair $(n^+,n^-)$
of mutually inverse bijections
as above,  commuting with {\it all} internal symmetries of the universe
$(\XX^+,\XX^-)$;
we then say that our geometry is {\it of the first kind}, or
 a {\it null geometry}. 
 In this case, the identification $\XX^+ \cong \XX^-$
is much more canonical than for any Hamilton operator, so that it
makes indeed sense to call the point $n^-(\alpha)$ for
an observer $\alpha$ its ``point at infinity''.
An apparently trivial example of this situation is the projective line
$\XX^+ = \K \PP^1$, which is canonically the same as its dual projective
line $\XX^-$ of ``hyperplanes'' (=points) in $\K \PP^1$; 
but clearly this identification  
is a null-system (a line is never transversal to itself!)  and not a polarity
(algebraically, this identification comes from the canonical symplectic form on
$\K^2$). 
We are thus forced to switch constantly between two {\it different}
ways of identifying $\XX^+$ and $\XX^-$, namely by the Hamilton operator $p$,
who governs the geometry of the active world, and by the ``underlying''
null energy $n$. This is indeed a good reason for distinguishing
$\XX^+$ and $\XX^-$ from the outset.
Without this clear distinction,  one arrives at paradoxes such as
``every active observer is both identical with its zero vector and with
its point at infinity''. The observer might simply say ``I am my origin
and my point at infinity''. 

\ssk {\bf Example 1.} {\em
The projective line $\Gras_1(\K^2)$ is generalized by the Grassmannians
$\Gras_n(\K^{2n})$, or, in arbitrary dimension, by
 the ``type = cotype'' Grassmann geometry
 $(\Gras_E^E(W) , \Gras_E^E(W))$
with $W=E \oplus E$: the identity map $\XX^+ \to \XX^-$ is
indeed a canonical null-system. 
Note that in this case $\XX^+$ and $\XX^-$ are modelled on 
$\End(E)=\Hom(E,E)$, which
is an {\em associative algebra}. 
%

\ssk
The Lagrangian geometry introduced above is also a null geometry, where the
null system is the identity map from $\XX^+$ to $\XX^-$. 
Since quantum mechanics corresponds to the choice of a base point $(o,o')$
in this geometry, polarities should always be compatible with this base point,
i.e., $o'=p^+(o)$. The fixed Hilbert structure on $\HH$ corresponds to the
choice of the {\em standard elliptic polarity} given by orthocomplementation
in the Hilbert space $W=\HH \oplus \HH$, or to the {\em standard hyperbolic
polarity}, given by orthocomplementation with respect to
 the neutral form $\beta( (u,v),(u',v')) :=
\langle u,u' \rangle - \langle v,v' \rangle$. 
}

\ssk {\bf Example 2.} {\em The example of the projective line 
is indeed quite typical: it is not misleading to consider
 ``null geometries'' as a rather subtle generalisation of the projective line.
In fact, as mentioned above, $\End(E)$ is an associative algebra; so let us
start with a general associative
algebra $A$ and  define $\XX^+ = \XX^-$ to be the {\em projective line
over $A$}, which by definition (cf., e.g., \cite{Bue95}) is the set
$A\PP^1:=\Gras_A^A(A \oplus A)$ of all
submodules of the left $A$-module $A \oplus A$ that are isomorphic to $A$
and admit a complementary submodule isomorphic to $A$.
This geometry is modelled on $A$, and again the identity map $\XX^+ \to
\XX^-$ is a canonical null-system.

\ssk 
More important for physics are subgeometries of the projective line
that are induced by fixing an {\em involution} 
 $*:A \to A$ (antiautomorphism of order $2$; if the base field
$\K$ itself carries a distinguished involution, we may also require that $*$ is
antilinear with respect to this involution). Then the involution $*$
 lifts to an
involution of the projective line $A \PP^1$ whose fixed point set
is called the {\em Hermitian projective line}, cf.\ \cite{BeNe05}, 
Section 8.
Again, this is a null geometry, and it generalizes the Lagrangian
geometry from Example 1.2.3.
It is therefore the geometric object corresponding to the abstract
 $C^*$-algebra approach to quantum mechanics.

\ssk
Of course, quantum mechanics requires to work over the field $\C$
of complex numbers and the involution $*$ to be $\C$-antilinear.
This has the particular consequence that the spaces of
Hermitian elements ($a^*=a$) and skew-Hermitian elements ($a^*=-a$)
are isomorphic, whereas for more general involutions this need not
be the case (e.g., real square matrices with $*$ being the usual
transpose). Correspondingly, in the general case there is also
a {\em skew-Hermitian projective line}, which in general is not
isomorphic to the Hermitian one (to be a bit more precise: the
skew-Hermitian projective line essentially 
corresponds to the unitary group of $(A,*)$, considered as a
homogeneous space under an even bigger group), but in the case of
quantum mechanics happens to be isomorphic to the Hermitian projective line.
This ``accident'' corresponds to the ambiguity in the 
interpretation of observables (see the quotation from \cite{AlSch01}
given in the Introduction), and we have the impression that it is
also related to the problem of time development
(``unitary $\bf U$-evolution versus
state reduction $\bf R$''; see item (4) in the Introduction and Subsection 
1.3.3.2 below).
}

\msk \nin
{\bf 1.3.3 Dynamics.} --
We affirmed above that a polarity creates dynamics.
This needs explanation: as a general fact, any polarity of a generalized
projective geometry defines, on the ``active universe'' $\MM^{(p)}$,
the structure of a {\em symmetric space} -- thus there is a
canonical torsionfree {\em affine connection} together with its
associated groups, and the notions of {\it geodesics}
and of {\it geodesic flow} on the tangent bundle $T \MM^{(p)}$ are defined.
This is best seen by looking at some examples, at least in the finite
dimensional real case; in the general infinite-dimensional case, these notions
are somewhat less standard, and we comment on this below. 

\ssk {\bf Examples.} {\em Let us first look at elliptic polarities of
finite-dimensional geometries, which lead to {\em compact symmetric spaces}:
in all cases, this polarity is given by orthocomplementation with respect to
a (real or complex) scalar product, so the relevant automorphism groups
are orthogonal, resp.\ unitary groups, which act transitively on the geometry.
We thus can write $\XX^+ \cong U/K$, $U$ a compact Lie group and $K$ 
essentially the group of fixed points of an involution $\sigma$ of $U$:

\begin{itemize}
\item for the real Grassmannians $\XX^+ \cong \OO(p+q)/\OO(p) \times \OO(q)$,
\item for the complex Grassmannians $\XX^+ \cong \UU(p+q)/\UU(p) \times \UU(q)$
(the special case $p=1$ corresponds to ordinary projective spaces),
\item  for the complex Hermitian
 Lagrangian geometry $\XX^+ \cong \UU(n) \times \UU(n)/diag \cong
\UU(n)$.
\end{itemize}

\nin 
These spaces are well known to be compact symmetric spaces, and the last two
series are moreover {\em Hermitian symmetric spaces} (\cite{Hel78}, 
\cite{Lo69}). It is known that all compact symmetric spaces of ``classical
type'' can be obtained in a similar way (cf.\ \cite{Be00}).
 The {\em non-compact dual} $G/K$ of $U/K$ can be obtained by taking a
slightly modified polarity (hyperbolic polarity);
in this case, the active universe
$M^{(p)}$ is a dense open subset of $\XX^+$ which
 is no longer connected, but one
of the connected components is always the non-compact dual $G/K$
(the inclusion $G/K \subset \XX^+ = U/K$ generalizes the well-known
{\em Borel-imbedding}, cf.\ \cite{Hel78}).
For instance, the non-compact duals of the projective spaces
are the {\em real, resp.\ complex hyperbolic spaces}, realized as ``balls''
in $\R^n$, resp.\ $\C^n$ (cf.\ also \cite{Fr05}).
Finally,
 it is also possible to choose polarities that are of yet
 different type, such that
$\MM^{(p)}$ is a {\em pseudo-Riemannian} symmetric space. For instance,
the de Sitter and anti-de Sitter models of general relativity are obtained
by suitable polarities of the Lagrangian geometry for $n=2$.
Replacing $\C^n$ by a Hilbert space and the groups $\UU(p,q)$
by suitable infinite dimensional unitary groups, these examples generalize.
In particular, the elliptic polarity of the infinite dimensional Lagrangian
geometry leads to the identification of the
``conformal completion $\XX^+$ of $\Herm(\HH)$'' with the unitary group
$\UU(\HH)$, which  here is seen as an
 infinite dimensional
Hermitian symmetric space $\UU(\HH) \times \UU(\HH)/diag \cong \UU(\HH)$.
}

\msk \nin
Concerning the dynamical system induced by a polarity, there are (at least)
two serious problems:

\msk \nin {\bf 1.3.3.1 Integration.} --  The first problem is of
analytic nature. It arises 
when we try to ``integrate'' the differential equations of the dynamical system,
for instance, to  obtain the geodesic flow corresponding to a spray. Here, 
properties of the base ring
$\K$ start to play a r\^ole: whereas the choice of $\K=\R$ or $\K=\C$
implies a tendency towards
 a more deterministic behavior (via existence and uniqueness theorems
for ordinary differential equations),
this may be much less the case for other base fields or rings
(for instance, there are no general existence and
 uniqueness theorems for $p$-adic
differential equations). For physics this may seem not to be a too serious
issue since one is mainly interested in real or complex
 Hilbert or Banach space settings and
not in ``too wild'' infinite dimensional situations where basic theorems
on ordinary differential equations may fail.

\msk \nin {\bf 1.3.3.2 The $\bf U$-evolution.} -- 
The second problem is of a more 
geometric nature. It arises when we try to identify, in the case of
the Lagrangian geometry corresponding to usual quantum mechanics, 
the  {\em $\bf U$-evolution} (unitary evolution defined by
the Schr\"odinger equation) with a flow defined in terms of geometry
(say, with the geodesic flow of a symmetric space).
If a base point $(o,o')$, an additional observable $H$ and a polarity $p$
are fixed, there are several possibilities to associate dynamical systems
to this situation, but at present none of them really seems to coincide with
the usual $\bf U$-evolution. Perhaps should it be necessary to use the description
by an observer
$ \alpha = \alpha(t)$ which also moves in $\XX^-$, in order to obtain the usual
picture? Or is it indeed necessary to invoke
 some ``associative feature'' that cannot be captured by a
Jordan description alone (cf.\ the quotation from \cite{AlSch01})? 
In \cite{AS98} it is explained, following Kibble
\cite{Ki79}, that the Schr\"odinger evolution can indeed be interpreted as
a Hamiltonian flow in a suitable geometric
setting; thus, at least, it seems reasonable
to look for an interpretation of the $\bf U$-evolution in geometric terms also in
the present setting.

\msk \nin
{\bf 1.3.4 Curvature.} -- 
Regardless
whether we are able to integrate differential equations  or not,
there is always a notion of {\it curvature} of an affine connection -- 
in general, the symmetric spaces defined by a polarity will
have non-vanishing curvature. The curvature tensor is a trilinear map
satisying the defining identities of a {\em Lie triple system} (see
Section 2.6), and it contains all local
information about the symmetric space, just as does the Lie algebra of
a Lie group. The Jordan product is closely related to the Lie triple
system (Section 2.6), and thus the Jordan product can geometrically
be interpreted as a  curvature feature.

\msk \nin
{\bf 1.4 States and pure states.} --
We started with observables (and observers) as fundamental objects of our
theory, and not with {\em states}.
On the other hand, in usual quantum mechanics, the pure states form
a projective space $\PP (\HH)$, and it might seem more natural to
take the ``geometry of pure states'' $(\PP (\HH),\PP(\HH^*))$ as basic object
of a geometric approach to quantum mechanics -- this is indeed the common
ground of all such approaches we know about,  see, e.g.,
 \cite{AS98}, \cite{BH01}, \cite{CGM03}, \cite{Ki79}, \cite{Va85}.
What is the relation between these approaches and the one proposed here?

\ssk
In a linear pair geometry $(\XX^+,\XX^-)$ there is a natural notion
of {\em intrinsic subspace} or {\em state (in $\XX^+$)}: it is defined as
a subset $\II \subset \XX^+$
 which, to {\em any} observer $\alpha \in \XX^-$, appears linearly, i.e.,
$\II \cap \alpha^\top$ is a linear subspace of $\alpha^\top$,
regardless which origin $o \in \II \cap \alpha^\top$ we choose, and it is
called {\it minimal} or a {\it pure state} if it is of {\em rank $1$}, i.e.,
it is not reduced to a point and does not properly
 include  intrinsic subspaces that are not points. 
Similarly, {\em states in $\XX^-$} are defined; we may call them
``dual states'' or ``kets''.
The best way to get some idea on these concepts is to look at examples:

\ssk {\bf Example 1.} {\em States of the Grassmann geometry
$(\XX^+,\XX^-)=(\Gras_E^F(W),\Gras_F^E(W))$ can be constructed as follows:
fix some flag $0 \subset F_1 \subset F_2 \subset W$ and let
$$
\II := \{ A \in \XX^+ | \, F_1 \subset A \subset F_2 \}
$$
be the set of all elements of the Grassmannian
 that are ``squeezed'' by this flag.
Then $\II$ is an intrinsic subspace. Conversely, 
 if $\K$ is a field and $W = \K^n$
is finite-dimensional, then all  intrinsic subspaces are of this form
(\cite{BL06}, Theorem 3.11). 
Such a state is pure if the codimensions are minimal, i.e.,
if $\dim E = \dim F_1 + 1 = \dim F_2 - 1$. 
In particular,
in the
 case of an ordinary projective geometry $\PP(W)$ (i.e., $\dim E=1$), states
are the usual projective subspaces in $\PP(W)$, 
and pure states are projective lines in $\PP(W)$.
As is well-known, every affine subspace 
in the affine picture then corresponds to a  state
(namely, to the projective subspace which is its completion).
 The situation changes drastically if $\dim E > 1$:
then only rather specific affine subspaces of the affine part belong
to states, namely the so-called {\em inner ideals} of the corresponding
Jordan pair (see Subsection 2.7).  
In particular, it follows that pure states through the origin
of the affine part $\Hom(E,F)$ are represented by rank-one operators in
the usual sense. }

\ssk {\bf Example 2.} {\em States of the Lagrangian geometry are
constructed similarly, by taking {\em Lagrangian flags} (i.e., flags 
$0 \subset F_1 \subset F_2 \subset W$ with $F_1^\perp = F_2$).
Then similar results as in the Grassmann case hold; in particular,
pure stats through the origin of the linear part $\Herm(\HH)$ are
represented by Hermitian rank-one operators, i.e., by projectors
on one-dimenional subspaces of $\HH$. We thus recover the space
of pure states of usual quantum mechanics: it is the space of all
minimal intrinsic subspaces running through a fixed ``origin''
$o \in \XX^+$. }

\msk \nin {\bf 1.4.1 The geometry of states.} --
 We denote by $\SSSS^\pm$ the
collection of all states in $\XX^\pm$, or perhaps of all pure states,
or of states of some given rank $r$,
and we think of  states as elements of a new
geometry $(\SSSS^+,\SSSS^-)$ which is  associated to our universe
$(\XX^+,\XX^-)$ in a similar way as, for example, one associates
to a usual pair of dual projective spaces $(\PP V,\PP V^*)$  the geometry of 
{\it all} projective subspaces, or of subspaces of a given rank.
One may ask whether $(\SSSS^+,\SSSS^-)$ is again a ``good'' geometry --
a look at our standard examples shows that this is indeed quite often
the case:

\ssk {\bf Example.} {\em
The preceding Example 1 shows that intrinsic subspaces of Grassmannians
correspond to {\em short flags} $0 \subset F_1 \subset F_2 \subset W$,
where the rank of the intrinsic subspace corresponds to the {\em type of
the flag} (characteristic dimensions, for instance). Thus
$(\SSSS^+,\SSSS^-)$ is a flag geometry. As mentioned in the example of
Section 1.2, such geometries are again linear pair geometries.
However, they are in general no longer affine pair geometries:
indeed, they are of the form $(G/P^-,G/P^+)$, where the parabolic subgroups
$P^\pm$ are no longer associated to $3$-gradings, but to $5$-gradings
(see Section 2.1). Such geometries attract much interest in current
research since they are related to ``non-commutative Jordan structures''
and also to exceptional geometries. 
The same remarks hold for the geometry of states of a
Lagrangian geometry: it corresponds to certain $5$-gradings of the Lie algebra
of ${\rm SU}(n,n)$. 

\ssk
Coming back to the geometry of quantum mechanics, we now have to
carefully distinguish between two notions of ``pure states'': first, the
usual one (minimal intrinsic subspaces running through the 
fixed origin, leading to projective space $\PP(\HH)$), and second,
the new base point-free interpretation (leading to the geometry of short
Lagrangian flags). In view of the preceding remarks, this change
corresponds to passing from usual, commutative Jordan algebraic structures
to non-commutative ones.     Thus we enter into a
``non-commutative Jordan geometry'' --
if the term ``second quantization'' were not already taken, it would be
tempting to use it here. 
Of course, we do not know if this new non-commutativity is reflected 
anywhere in physical reality, but at least theoretically there might
be some possibility here to test  experimentally whether the second
interpretation has some physical meaning. }

%

\msk \nin {\bf 1.4.2 Comparison with twistor geometry.} -- 
Another striking feature of our interpretation of states is the similarity
with the double fibrations from Penrose's twistor theory:
the interpretation of states as intrinsic subspaces
produces a new duality --  
a duality between the ``geometry of observables'' $(\XX^+,\XX^-)$
 and the ``geometry of states'' $(\SSSS^+,\SSSS^-)$. This new duality
corresponds to  the double fibration 
$$
\begin{matrix}   & \ZZ^+  & \\
 & \swarrow \quad \quad  \quad \quad  \searrow & \\
\XX^+ & & \SSSS^+ \\
\end{matrix}
$$
where
$\ZZ^+ = \{ (x,\II)  | x \in \II \} \subset  \XX^+ \times \SSSS^+ $
is the ``incidence space'' (``$x$ is incident with $\II$'').

\ssk {\bf Example.} {\em
For $\K=\C$ and $\XX^+ =  \Gras_2(\C^{4})$,
$\SSSS^+ = \C \PP^3 = \Gras_1(\C^4)$, we get the complexified setting of
Penrose's twistor theory as described in \cite{BE89}, p.\ 8.
Here, $\SSSS^+$ rather corresponds to a geometry of {\em maximal}
 intrinsic subspaces.
In the more sophisticated setting from \cite{Pen05}, Chapter 33, starting with
(compactified)
{\em real} Minkowski space, light rays indeed correspond to minimal intrinsic 
subspaces, and $\SSSS^+$ forms  a $5$-graded geometry of Lagrangian flags
associated to 
${\rm SU}(2,2)$. 
%
In any case, our setting incorporates  aspects of {\em non-locality} that
have been a main motivation of Penrose's for developping twistor theory:
a pure state is a ``line'' and thus is a {\em global} object of our universe
$\XX^+$.
Note finally the  change of r\^ole of ``observables'' and ``states'':
by analogy with quantum theory, we are driven to call ``pure state''
the light rays and not the points of Minkowski space, which rather correspond to
``observables'', whereas classically points of a manifold are rather viewed
as pure states.
Compare with \cite{Pen05}, p.\ 964: 
``there is a striking reversal of this in twistor space, since now the
light ray is described as a {\em point} and an event is described as
a {\em locus}.''
%
}

\msk
\nin {\bf 1.5 Open ends.} --
The reader who
wishes to recover the familiar picture of quantum mechanics will 
wonder (at least) about the following questions:

\ssk (1) How is ``measurement'' and ``state reduction'' described?
As to the mathematical framework, the analog of ``eigenvectors'' and
``diagonalization'' (with respect to some complete family of pure states
through a given point, which is the analog of a Hilbert basis) is perfectly
well-defined (I will not go into details; the specialist may look at \cite{Lo91}). 
As to its interpretation as ``state reduction'', I prefer not
to make any statements -- following the advice of J.\ Bell (\cite{Bell04}, p.\
126): ``Concepts of `measurement', or `observation', or `experiment' should
not appear at a fundamental level.'' 

\ssk (2) 
What, after all, shall be the interpretation of the
 ``universe'' $(\XX^+,\XX^-)$:
 shall we think of it  as space-time of relativity,
or quantum mechanically as
a {\em single} particle, or as a {\em many-particle system}, 
or as a field of ``all''
particles? According to the standard formalism of quantum mechanics, there should
be some way to compose a many-particle system from single particles,
formalized by the tensor product of Hilbert spaces; but there is no
``tensor product of Jordan algebras'', and accordingly there is no obvious
notion of ``tensor product of generalized projective geometries''.
This is indeed a serious problem; I will comment in the last chapter
 on some possible ways of attack.

\section{Jordan theory}

\nin
{\bf 2.1 Historical remark.} --
As mentioned in the Introduction,
 Pascual Jordan's Ansatz started with
the observation that any associative algebra, such as finite or 
infinite-dimen\-sio\-nal matrix algebras, equipped with the new product
$x \bullet y = \frac{1}{2} (xy+yx)$, becomes a commutative algebra which is
not associative but satisfies another identity, namely
$$
(x^2 \bullet y) \bullet x = x^2 \bullet (y \bullet x).
$$
In case that $2$ is invertible in $\K$ (which we will always assume),
this identity then served
as axiomatic definition of a class of commutative algebras, the
later so-called {\em Jordan algebras} (see Part I of \cite{McC04} for a 
detailed historical survey of Jordan theory).
Unfortunately, this axiomatic definition is much less appealing than
the one of a Lie algebra, and therefore we prefer to skip some 40 years
of historical development and turn right away to more general objects
that are easier understood, namely to {\it Jordan pairs} and
{\em Jordan triple systems}. Let us, however, point out that a Jordan algebra
is called {\em special} if it is a subalgebra of some associative algebra
with the symmetrized bullet-product; there exists essentially only one
exceptional Jordan algebra (the 27-dimensional {\em Albert algebra}, which
has attracted some attention also in physics, most notably by work of 
M.\ G\"unaydin and his collaborators,
see the extensive bibliography
in \cite{Io03}.) The other 
classical Jordan algebras are:

\msk  {\bf Example.} {\em Here are the main families of special Jordan algebras:
\begin{itemize}
\item[\rm (1)] full matrix algebras $M(n,n;\K)$ (full endomorphism algebras of a
linear space),
\item[\rm (2)] symmetric and Hermitian matrices $\Sym(n,\K)$, resp.\ $\Herm(n,\K)$
(selfadjoint operators of an inner product space),
\item[\rm (3)] skew-symmetric matrices in even dimension (selfadjoint operators with
respect to a symplectic form),
\item[\rm (4)] spin factors: vector spaces $V$ with symmetric
bilinear form $\beta:V \times V
\to \K$ and a distinguished element $e$ with $\beta(e,e)=1$, with product
$$
x \bullet y := \beta(x,e)y + \beta(y,e)x - \beta(x,y)e.
$$
\end{itemize}
}

 \nin {\bf 2.2 Jordan pairs and $3$-graded Lie algebras.} --
A {\em $2k+1$-graded Lie algebra} is a Lie algebra of the form 
$$
\g = \g_{-k} \oplus \ldots \oplus \g_0 \oplus \g_1 \oplus \ldots \oplus \g_k
$$ 
($k \in \N$) such that
$[\g_i,\g_j] \subset \g_{i+j}$.
By convention, $\g_i = 0$ for $i \notin \{ -k,\ldots,k \}$.
We are particularly interested in $3$-graded Lie algebras ($k=1$); then 
our the condition implies
that $\g_1$ and $\g_{-1}$ are abelian subalgebras of $\g$. 
 Let $V^\pm := \g_{\pm 1}$.
We define trilinear products
$$
T^\pm : V^\pm \times V^\mp \times V^\pm \to V^\pm, \quad (x,y,z) \mapsto
[[x,y],z].
$$
From the Jacobi identity in $\g$, together with the fact that $\g_{\pm 1}$
are abelian subalgebras of $\g$, we then easily get the following two 
identities:

\begin{description}
\item{(LJP1)} $T^\pm(x,y,z) = T^\pm(z,y,x)$,
\item{(LJP2)} $T^\pm(a,b,T^\pm(x,y,z)) =$
\item{ } $\quad \quad \quad \quad \quad T^\pm(T^\pm(a,b,x),y,z) -
T^\pm(x,T^\mp(b,a,y,z)) + T^\pm(x,y,T^\pm(a,b,z))$.
\end{description}

\nin By definition, a {\em (linear) Jordan pair} is a pair $(V^+,V^-)$
of $\K$-modules together with trilinear maps
$T^\pm : V^\pm \times V^\mp \times V^\pm \to V^\pm$ satisfying (LJP1)
and (LJP2).
Every linear Jordan pair is obtained by the construction just described
(sometimes this is called the {\em Kantor-Koecher-Tits construction};
the easiest proof is by noting that $V^+ \oplus V^-$ carries the structure
of a (polarized) {\em Lie triple system}, and then the so-called
{\em standard imbedding}
of this Lie triple system yields a $3$-graded Lie algebra (cf.\ \cite{Be00},
Chapter III).

\ssk {\bf Example.} {\em All classical Lie algebras admit $3$-gradings
(if we allow $\mathfrak{sl}(n,\K)$ to be replaced by $\gL(n,\K)$).
Let us consider the Lie algebra $\g=\gL(W)$ of all endomorphisms of a linear
space $W$ and fix a direct sum decomposition 
 $W=E \oplus F$. Then we have an associated $3$-grading
$\g=\g_{-1} \oplus \g_0 \oplus \g_0$, where, in an obvious matrix notation,
$$
\g_1 := \Big\{ \begin{pmatrix} 1 & x \cr 0 & 1 \cr \end{pmatrix} | \, x \in
\Hom(F,E) \Big\}, \quad \quad
\g_{-1} := \Big\{ \begin{pmatrix} 1 & 0 \cr y & 1 \cr \end{pmatrix} | \,
y \in \Hom(F,E) \Big\} ,
$$
and $\g_0$ given by the ``diagonal matrices'' in $\g$.
Calculating the triple bracket $[[x,y],z]$ for $x,z \in \g_1$, $y \in \g_{-1}$,
we see that the corresponding Jordan pair is 
$$
(V^+,V^-) = (\Hom(F,E),\Hom(E,F)) \quad \mbox{with} \quad
T^\pm(x,y,z)=xyz+zyx.
$$
Next,
the symplectic Lie algebra $\sP(\HH\oplus \HH)$ and the orthogonal Lie
algebra $\oo(\HH \ominus \HH)$, with respect to the symplectic form
$\omega((u,v),(u',v')) = \langle u,v' \rangle - \langle v,u' \rangle$,
resp.\ with respect to the symmetric form
$\beta ((u,v),(u',v')) = \langle u,v' \rangle + \langle v,u' \rangle$
on $\HH \oplus \HH$, are subalgebras of $\gL(\HH \oplus \HH)$
and inherit from it a $3$-grading.
The corresponding Jordan pairs are of the form $(V^+,V^-)=(V,V)$ where
$V$ is the space of symmetric (or Hermitian), resp.\ of skew-symmetric (or 
skew-Hermitian) matrices, with trilinear map given by the same formula as above.
The similarity in notation with the description of Grassmann and Lagrangian
geometries is of course not an accident (see explanations below).
Finally, some orthogonal Lie algebras  admit
another family of $3$-gradings, leading to Jordan pairs related to
the spin-factors defined above.
Exceptional Lie algebras not always admit a $3$-grading; for instance,
$G_2$, $F_4$ and $E_8$ do not, whereas some forms of $E_6$ and $E_7$ do.}

\msk \nin {\bf 2.3 Jordan triple systems and involutive
$3$-graded Lie algebras.} --
Assume now that $\g$ is a $3$-graded Lie algebra together with an
involution $\theta$, i.e., an automorphism of order $2$ reversing the
grading: $\theta(\g_i)=\g_{-i}$.
Then let $V:=V^+ = \theta(V^-)$ be equipped with the trilinear product
$$
T(x,y,z):= T^+(x,\theta y,z) = [[x,\theta y],z]
$$
which clearly satisfies the two identities (JT1) and (JT2) obtained from
(LJP1) and (LJP2) by omitting the
superscripts $\pm 1$. 
By definition, a {\em Jordan triple system} is a $\K$-module $V$ together
with a trilinear map $T:V \times V \times V \to V$ satisfying 
(JT1) and (JT2). 
Every Jordan triple system is obtained by the construction just described:
from $(V,T)$ one recovers a Jordan pair $(V^+,V^-,T^\pm)$ by letting
$V^+ := V^- :=V$ and $T^\pm:=T$, and then applies the ``Kantor-Koecher-Tits
construction'' outlined above.

\ssk  {\bf Example.} {\em All $3$-graded Lie algebras from the
preceding example admit involutions, but in general there is no distinguished one.
The matrix realization given above priviliges the involution given by
$\theta(X)=-X^t$ (negative transpose), which leads to the matrix
triple systems given by $T(X,Y,Z)=-(XY^tZ+ZY^tX)$.}

\msk \nin {\bf 2.4 Unital Jordan algebras and invertible elements.} --
Let $\g$ be a $3$-graded Lie algebra, with corresponding Jordan pair
$(V^+,V^-)$.
For $x \in V^\pm$, the {\em quadratic operator} is defined by
$$
Q^\pm(x):V^\mp \to V^\pm, \quad y \mapsto \mathfrak{1 \over 2} T^\pm(x,y,x)=
 \mathfrak{1 \over 2} [[x,y],x].
$$
An element $x \in V^-$ is called {\it invertible} if the linear map
$Q^-(x):V^+ \to V^-$ is invertible. It can be shown that then
 $V:=V^+$ with the bilinear product
$$
y \bullet z :=  \mathfrak{1 \over 2} 
T^+(y,x,z)
$$
becomes a Jordan algebra with unit element $e:=Q(x)^{-1}(x)$,
and that every unital Jordan algebra is obtained in this way.
In other words, unital Jordan algebras are the same as Jordan pairs
together with a distinguished invertible element.

\ssk {\bf Example.} {\em 
For $\gL(W)$ with $W=E \oplus F$, the quadratic operator is given by
$$
Q^+(x):\Hom(E,F) \to \Hom(F,E), \quad y \mapsto xyx.
$$
If $\K$ is a field and $E$ and $F$ are non-isomorphic vector spaces, then this
map cannot be an isomorphism, and hence the Jordan pair never has invertible
elements. On the other hand, if $E$ and $F$ are isomorphic as vector spaces,
 then $Q^+(x)$ is bijective if and only
if $x$ is invertible as a linear map from $E$ to $F$, 
and then $Q^+(x)^{-1}=Q^-(x^{-1})$.
Fixing for a moment such an element
$x$ as an identification of $E$ and $F$, the corresponding
Jordan algebra structure on $V :=V^+ \cong V^- \cong \End(E)$ is just the usual
symmetrized product.
Similarly, the Jordan pairs of symmetric and Hermitian matrices always have
invertible elements, whereas for $n\times n$-skew symmetric matrices this is true only
when $n$ is even. 

\ssk
Note that the bullet-product defined above depends on $x$. In the case of
full matrix pairs $(\End(E),\End(E))$, this dependence is not very serious:
as long as $x$ is invertible, all these products are isomorphic, but for
Hermitian matrix pairs $(\Herm(\HH),\Herm(\HH))$, it becomes more subtle:
it depends on the isomorphism class of $x$, seen as a Hermitian form on $\HH$;
thus in general  the various Jordan algebras obtained by
fixing the invertible element $x$ need no longer be isomorphic among each other
(they are only ``isotopic''). 
}

\msk \nin {\bf 2.5 The fundamental formula.} --
The following identity which is valid in all Jordan pairs $(V^+,V^-)$ is called the
{\em fundamental formula}: for all $x \in V^-$, $y \in V^+$,
$$
Q^-(Q^-(x)y) = Q^-(x)Q^+(y)Q^-(x).
$$
There seems to be no ``straightforward proof'' of this formula, starting from
the definition of a linear Jordan pair
 given in Subsection 2.2. In \cite{Lo75} this formula is taken as one of
the {\em defining} axioms of a Jordan pair.

\ssk {\bf Example.} {\em In case of the Jordan pair of rectangular matrices,
$(\Hom(E,F),\Hom(F,E))$, the proof of the fundamental formula is easy:
as seen above $Q(x)z=xzx$, and hence
$Q(Q(x)y)z=Q(xyx)z=xyxzxyx=Q(x)Q(y)Q(x)z$.}

\msk \nin {\bf 2.6 The Jordan-Lie functor.} --
To every Jordan triple system $(V,T)$ we associate a new ternary product 
$R=R_T$ by antisymmetrizing in the first two variables:
$$
[X,Y,Z]:= R_T(X,Y)Z := T(X,Y,Z) - T(Y,X,Z).
$$
It is easily shown that this trilinear product  is a {\em Lie triple system}, i.e.,

\begin{itemize}
\item it is antisymmetric in $X$ and $Y$, 
\item it satisfies the Jacobi identity 
$[X,Y,Z]+[Y,Z,X]+[Z,X,Y]=0$, 
\item the endomorphism $R_T(X,Y):V \to V$ is a {\it derivation} of the
triple product $[\cdot,\cdot,\cdot]$. 
\end{itemize}

\nin
The correspondence $T \mapsto R_T$ is called the {\em Jordan-Lie functor}
(see \cite{Be00}). 
Lie triple systems are for symmetric spaces what Lie algebras are for Lie groups:
an infinitesimal version that, locally, determines them completely. 
Moreover, the Lie triple product
is (possibly up to a sign) the {\em curvature tensor of the
canonical connection} (see \cite{Lo69}, \cite{Be06}).
Many, but not all Lie triple systems are obtained from Jordan triple systems
via the Jordan-Lie functor; and some Lie triple systems are obtained from
2 or even 3 different Jordan triple systems
(cf.\ tables in \cite{Be00}, Ch.\ XII).

\ssk {\bf Example.} {\em Consider the Jordan triple system $V^+ = V^- =
\End(E)$ with $T(x,y,z)=xyz+zyx$. In this case
$$
R_T(x,y)z= xyz + zyx - yxz - xxy = [x,y]z - z[x,y] = [[x,y],z]
$$
is the triple Lie bracket in $\gL(E)$. 
For other classical Lie algebras, the triple Lie bracket can be similarly
obtained from a Jordan triple system, but never for exceptional Lie algebras.
``Classical'' symmetric spaces essentially 
can all be realized as subspaces fixed under
one or several involutions (involutive automorphisms or anti-automorphisms)
 of $\Gl(n,\K)$, and hence their Lie triple system
comes from a Jordan sub-triple system of $M(n,n;\K)$ (see \cite{Be00});
for exceptional symmetric spaces the situation is more difficult.
}

\msk \nin {\bf 2.7 Inner ideals.} --
Left and right ideals of associative algebras are generalized by
{\em inner ideals} in Jordan theory: an {\em inner ideal} $I$ in the
$+$-part $V^+$ of a Jordan pair $(V^+,V^-)$ is a linear subspace $I \subset V^+$
which is stable under ``multiplication from the inside'':
$$
T^+(I,V^-,I) \subset I .
$$
John R.\ Faulkner has proposed to use inner ideals as a key ingredient for an
incidence geometric approach to Quantum Theory \cite{F80}.

\ssk
 {\bf Example.} {\em Consider the Jordan triple system $V:=V^+ = V^- =
\End(E)$ with $T(x,y,z)=xyz+zyx$, and let $L \subset \End(E)$ be a left ideal
in the usual sense. Then
$T^+(L,V^-,L) \subset L V L + L V L \subset L$; hence $L$ is an inner ideal.
Similarly for right ideals $R$. Clearly, intersections of inner ideals
are inner ideals; hence $L \cap R$ is an inner ideal. In finite dimension over
a field, all inner ideals of $V^+$ are of this form -- see \cite{BL06}, Appendix A
for an elementary account on this.}

\section{Generalized projective geometries} 

This chapter makes the link between the preceding two ones: we define
{\em generalized projective geometries} and explain the {\em Jordan
functor}: Jordan structures are for generalized projective (and polar) geometries
what Lie algebras are for Lie groups, namely an ``infinitesimal version''.
This correspondence works even better than in the Lie case: we can go
backwards and ``integrate'' Jordan structures to geometries, regardless
of the dimension and of the nature of the base field. Summing up, Jordan algebraic
structures and their corresponding geometries (equipped with a base point) are
completely equivalent. As before, $\K$ is a commutative unital base ring, such
 that $2$ and $3$
are invertible in $\K$. (The generalization of the following concepts
to the case of characteristic $2$ is an interesting open problem.)

\msk \nin {\bf 3.1 Some categorial notions for linear pair geometries.} --
Recall the notion of {\em linear} and {\em affine pair geometries} from Section 1.2.
 Although implicitly most has already been said in Chapter 1,
let us be more explicit about categorial notions for such
pair geometries, such as {\em morphisms}, {\em direct products and function spaces},
{\em connectedness}, {\em faithfulness}.

\msk \nin {\bf 3.1.1 Morphisms.\ I.} -- A
{\em homomorphism} between linear pair geometries $(\XX^+,\XX^-)$
and $(\YY^+,\YY^-)$ is a
 pair of maps $(g^+,g^-):(\XX^+,\XX^-) \to (\YY^+,\YY^-)$  
preserving transversality and being compatible
with the structure maps in the sense that
$$
g^+ \PPP_r(x,\alpha,y)=\PPP_r(g^+(x),g^-\alpha ,g^+(y)),
$$
 and dually.
This means simply that $g^+$ induces by restriction
a {\em linear} map from $(\alpha^\top,x)$ to $((g^-(\alpha))^\top,g^+(x))$, and
dually. In particular, we can speak of the {\it automorphism
group} $\Aut(\XX^+,\XX^-)$.
If a base point $(o^-,o^+)$ is fixed, then we call
 {\it structure group} the group
$\Aut(X^+,X^-;o^+,o^-)$ of automorphisms fixing the base point. 
From the definitions it follows that this group acts linearly
on the linear space $(o^-)^\top \times (o^+)^\top$.

\msk \nin {\bf 3.1.2  Morphisms.\ II.} --
{\em Adjoint} or {\em structural pairs of morphisms} are given by pairs 
$g:\XX^+ \to \YY^+$, $h:\YY^- \to \XX^-$
such that transversality is preserved in the sense that
$x \top h(\alpha)$  iff  $g(x) \top \alpha$, and, whenever
$(x,h(\alpha))$ and $(y,h(\alpha))$ are transversal, then
$$
g \PPP_r(x,h\alpha ,y) = \PPP_r(gx,\alpha ,gy)
$$
and similarly for $\PPP_r^-,\SSS^+$ and $\SSS^-$. We write $h=g^t$ if 
$(g,h)$ is a structural pair.
The condition means that $g$ induces a {\em linear}
 map from $((h\alpha)^\top,x)$ to $(\alpha^\top,gx)$. 
Note that every isomorphism $(g^+,g^-)$ in the sense of 3.1.1
 gives rise to a structural pair
$(g^+,(g^-)^{-1})$, and conversely, every bijective structural pair gives
rise to an isomorphism. 
Thus we have two different categories, but isomorphisms are 
essentially the same in both of them.
For more flexibility, we may consider, in both categories, pairs of maps that
are not necessarily defined everywhere.

\ssk {\bf Example.} {\em What is your preferred notion of a
 {\em homomorphism of projective
spaces}? Do you prefer  maps $\PP(W) \to \PP(V)$ that are induced by {\em injective}
linear maps $W \to V$ (hence are defined everywhere), or do you prefer maps
induced by arbitrary non-zero linear maps $W \to V$ (hence are not everywhere
defined as maps in the usual sense)?  
In the first case, you prefer to look at projective spaces as members of the
Category I (namely, in this category 
the projective geometry $(\PP(W),\PP(W^*)$ is a {\em simple} object,
 and hence homomorphisms have to be either injective or trivial).
In the second case, you prefer to look at them as members of the Category II
(namely, in this category, the dual map $f^*$ of an
arbitrary non-zero linear map $f:W \to V$ gives the pair $([f],[f]^t)=
[f],[f^*])$.)}

\msk \nin {\bf 3.1.3 Direct products and function geometries.} --
If $(\XX^+_i,\XX^-_i;\top_i)_{i \in I}$ is a family of linear or affine pair 
geometries, then the {\em direct product}, with transversality given by
$$
(x_i)_{i \in I} \top (\alpha_i)_{i \in I} \quad \mbox{iff} \quad
\forall i \in I:\, x_i \top \alpha_i
$$
is a linear (resp.\ affine) pair geometry: the new structure maps are simply
the direct products of those of $(\XX_i^+,\XX_i^-)$.

\ssk {\bf Examples.} {\em 
A particularly interesting special case is the direct product
of a geometry with its dual geometry,
$(\XX^+ \times \XX^-,\XX^- \times \XX^+)$. It  carries a canonical
polarity, namely the {\em exchange map} $\tau((x,\alpha))=(\alpha,x)$. 

\ssk Another important case are
{\em function geometries}: the index set is some geometric space $M$
and all $(\XX_i^+,\XX^-)$, $i \in M$, are copies of a fixed geometry $(\XX^+,
\XX^-)$. In other words, we consider the space of pairs of functions,
$$
(\Fun(M,\XX^+),\Fun(M,\XX^-)),
$$
equipped with the ``pointwise product''
$(\PPP_r(f,g,h))(x):= \PPP_r(f(x),g(x),h(x))$.
Specializing to the case $(\XX^+,\XX^-) = (\K \PP^1,\K \PP^1)$,
we get the geometric analog of the usual function spaces
$\Fun(M,\K)$. The philosophy of non-commutative geometry associates the
usual function spaces (which are commutative geometries) to Classical
Mechanics, whereas more general non-commutative, but still associative,
geometries are associated with Quantum Mechanics.
In some sense, the preceding constructions provide a non-associative
counterpart to this philosophy -- see \cite{Be07} for a further
discussion of this viewpoint.
}

\msk \nin {\bf 3.1.4
Faithfulness (non-degeneracy).} --
The geometry $(\XX^+,\XX^-)$ is called {\it faithful} if 
$\XX^-$ is faithfully represented by its effect of
 linearizing $\XX^+$, and vice versa: whenever 
 $\alpha^\top = \beta^\top$ as sets {\em and} as linear spaces
(with respect to some origin $o$), then $\alpha=\beta$, and dually.  
(Faithfulness corresponds to {\em non-degeneracy} in the Jordan-theoretic
sense.) Note that, in a faithful geometry, 
the component $g^+$ of an automorphism
$(g^+,g^-)$ determines uniquely the second component $g^-$ which must
correspond to push-forward of linear structures via $g^+$.

\ssk {\bf Example.} {\em The classical geometries (Grassmann or
Lagrangian geometries) are all faithful: the projective group is faithfully
represented by its action on $\XX^+$. 
A very degenerate geometry is the {\em trivial geometry}: take a pair
$(E,F)$ of $\K$-modules with trivial transversality relation
(all $(x,\alpha)$ are transversal), and all affine
structures are the same, equal to the given ones on $E$, resp.\ $F$.
}

\msk \nin {\bf 3.1.5
Connectedness and stability.} --
We will say that two points $x,y \in \XX^+$ are
{\em on a common chart} if there is $\alpha \in \XX^-$,
such that $x,y \in \alpha^\top$. 
Equivalently,  $x^\top \cap y^\top \not= \emptyset$. 
We will say that $x,y \in \XX^+$ are {\it connected}
if there is a sequence of points $x_0=x,x_1,\ldots,x_k=y$ such that
$x_i$ and $x_{i+1}$ are on a common chart.
This defines an equivalence relation on $\XX^+$ whose
equivalence classes are called {\it connected
components of $\XX^+$}. By duality, connected components
of $\XX^-$ are also defined. The geometry is called
{\it connected} if both $\XX^+$ and $\XX^-$ are connected.

\ssk
The pair geometry $(\XX^+,\XX^-,\top)$ will be called {\it stable}
if any two points $x,y \in \XX^+$ are on a common chart,
and dually for any pair of points $\alpha,\beta \in \XX^-$.
Clearly, a stable geometry is connected (the converse is not true).

\ssk {\bf Example.} {\em
In finite dimension over a field, Grassmann and symplectic
 Lagrange geometries are stable
because the affine parts $\alpha^\top$ then are Zariski-dense in
$\XX^+$, hence have non-empty
intersection. By contrast, the {\em total Grassmann geometry}
($\XX^+ = \XX^- =$ the set off {\em all} linear subspaces)
is in general highly non-connected (in finite dimension, its connected
components are the Grassmannians of subspaces of a fixed dimension).
}

\msk \nin {\bf 3.2 Laws.} --
We now describe some more specific laws which may or may not hold in
a linear pair geometry. We assume from now on that our geometry is an
{\it affine} pair geometry; the suitable formulation of laws for geometries
that are not affine is a completely open problem for the time being.
There are two ``fundamental laws of projective geometry'', denoted in the
sequel by (PG1) and (PG2), which can be put in a very consise form as
follows and on which we will comment in the sequel:
$$
 (L^{(r)}_{o,\alpha})^t  =  L^{(r)}_{\alpha,o} 
\eqno  \mbox{(PG1)}  
$$
$$
(M^{(r)}_{x,y})^t  =  M^{(r)}_{y,x} 
\eqno  \mbox{(PG2)}
$$
The notation $L,R,M$ refers to operators of
``left'', ``right'' and ``middle translations'': the ternary
map $\PPP_r$ gives rise to the operators
$$
\PPP_r(x,\alpha,y)=: L^{(r)}_{x,\alpha}(y) =:
R^{(r)}_{\alpha,y}(x) =: M^{(r)}_{x,y}(\alpha).
$$
Here, $L^{(r)}_{x,\alpha}$ is just the {\em dilation} denoted before by 
$r_{x,\alpha}$, and in an {\em affine} pair geometry we have the
simple relation $R^{(1-r)}_{\alpha,y} = L^{(r)}_{y,\alpha}$.
Whereas these two operators act on (parts of) $\XX^+$, the middle translation
$M_{x,y}^{(r)}$ acts from (a part of) $\XX^-$ to its ``dual'' $\XX^+$ !

\msk \nin {\bf 3.2.1 The First Law.} --
We say that an affine pair geometry {\em satisfies the First Law}
if (PG1) holds: for all transversal pairs
$(o,\alpha)$ and scalars $r \in \K$,
 the pair of
dilations (i.e., left translations)
$(r_{o,\alpha},r_{\alpha,o}) = (L^{(r)}_{o,\alpha},L^{(r)}_{o,\alpha})$
 forms a structural pair of morphisms. 
In view of the definition of a
structural pair  in Section  3.1.2, this can be written as an identity in $5$
variables: for all $r,s \in \K$,
$$
\PPP_r^+(o,\alpha,\PPP_s^+(x,\PPP_r^-(\alpha,o,\beta),y))
=  \PPP_s^+(\PPP_r^-(o,\alpha,x),\beta,\PPP^+_r(o,\alpha,y)).
\eqno  \mbox{(PG1)}  
$$
If $r$ is invertible, then the dilation $r_{o,\alpha}$ is invertible
on $\alpha^\top$. We require that the pair $(r_{o,\alpha},
r_{\alpha,o})$ extends to a bijection of $(\XX^+,\XX^-)$; then the identity
(PG1) can be interpreted by saying that the pair
$$
(g,g') := (r_{o,\alpha}, r_{\alpha,o}^{-1})
$$
is an automorphism (called an {\em inner automorphism}) of the affine pair
geometry $(\XX^+,\XX^-)$.
In particular, this means that to every transversal pair $(o,\alpha)$,
there is attached a homomorphic image of $\K^\times$, acting by automorphisms
of $(\XX^+,\XX^-)$ and preserving the pair $(o,\alpha)$.

The First Law thus ensures a rich supply of automorphisms, and
it is not hard to show that  {\em connected}
(PG1)-geometries are homogeneous: the automorphism group
$G=\Aut(\XX^+,\XX^-)$ acts transitively on $\XX^+$, on $\XX^-$ and
on $(\XX^+ \times \XX^-)^\top$, so that,
 with respect to a fixed base point 
$(o^+,o^-)$, we can write
$$
\XX^+ = G/P^-, \quad \quad
\XX^- = G/P^+, \quad \quad
(\XX^+ \times \XX^-)^\top = G/H \, \, \mbox{with} \, \, H = P^+ \cap P^-.
$$
In the non-degenerate
finite-dimensional cases over $\K=\C$ or $\R$, $G$ turns out to be
 a Lie group, and the groups
$P^\pm$ are maximal parabolic subgroups of $G$.

\ssk {\bf Example.} {\em The proof that the first law holds, 
for invertible scalars $r$,
is easy in the case of Grassmann and Lagrangian geometries: it suffices to remark first
that the natural action of the projective group $\PP \Gl(W)$ on 
$(\Gras_E^F(W),\Gras_F^E(W))$ is by automorphisms, and second that
the automorphism $r \id_E \oplus r^{-1} \id_F$ acts on 
the space of complements of $E$ by $r^{-1}$ and on those of $F$ by $r$; put together,
this is precisely (PG1). For non-invertible $r$, the proof is essentially the
same (cf.\ \cite{Be04}). 
Lagrangian geometries inherit this law since they are subgeometries of Grassmann
geometries.

\ssk
The First Law should be seen as the ``geometric version'' of the identity (LJP2)
of a Jordan pair: both identities have the same formal
 structure, with automorphisms and
products replaced by
derivations and sums,
 and the inversion of the scalar by the minus-sign in the middle term
in (LJP2).}

\msk \nin {\bf 3.2.1 The Second Law.} --
We say that an affine pair geometry {\em satisfies the Second Law}
if (PG2) holds, i.e., if, for all  pairs
$(x,y) \in \XX^+ \times \XX^+$ lying on some 
common chart $\alpha$
(i.e., $\alpha \in x^\top \cap y^\top$), and for all $r,s \in \K$,
the pair of ``middle translations'' $(M^{(r)}_{x,y},
M^{(r)}_{y,x})$ acts as a structural pair between the geometry $(\XX^+,\XX^-)$
and its dual geometry $(\XX^-,\XX^+)$.
As for the First Law, this is an identity in $5$ variables:
$$
\PPP_r^+(x,\PPP_s^-(\alpha,\PPP_r^+(y,\beta,x),\gamma),y)
 =  \PPP_s^+(\PPP_r^-(x,\alpha,y),\beta,\PPP^+_r(x,\gamma,y)).
\eqno  \mbox{(PG2)}  
$$
Of course, we require also the ``dual version'' of this identity to hold.
The scalar $r=\frac{1}{2}$ plays a special r\^ole with respect to this law,
because it satisfies
$M^{(r)}_{x,y} = M^{(r)}_{y,x}$, and hence the operator
$f=M^{(r)}_{x,y}$ is ``self-adjoint'' in the sense that $f^t = f$.

\ssk {\bf Example.} {\em The proof that the second law holds in a Grassmann geometry
is elementary, but considerably more tricky than the proof of the First Law --
one may use the explicit formula from Subsection 1.2.2 for the multiplication
maps and then prove (PG2) in a suitable affinization, see \cite{Be04}. 
In the course of that proof, one sees that (PG2) 
 is indeed the geometric analog of the Fundamental
Formula.}

\msk \nin {\bf 3.3 Generalized projective geometries.} --
A {\em generalized projective geometry over $\K$} 
is an affine pair geometry over $\K$ such that the First and the Second Law
are satisfied {\em in all scalar extensions of $\K$}.
The latter, slightly technical, condition is natural from an algebraist's point
of view; it ensures that to every geometry $(\XX^+,\XX^-)$ one can associate,
by scalar extension from $\K$ to the ring $\K[\eps]$ of dual numbers over
$\K$, in a functorial way, a new geometry $(T\XX^+,T\XX^-)$, defined over
$\K[\eps]$, and called the {\em tangent geometry} (cf.\  \cite{Be02}). 
In other words, one may
apply some algebraic version of differential calculus.
We advise the reader to think for the moment of $(\XX^+,\XX^-)$ as some
kind of smooth manifold, so that the tangent bundle in the usual sense existes;
by the usual chain rule of differential calculus, the tangent maps
$T \PPP$ then satisfy the same identies as the structure maps themselves,
so that the tangent geometry satisfies again (PG1) and (PG2) over the tangent
 ring $T\K$ which is nothing but $\K[\eps]$ (see \cite{Be06}
 for a justification
of this point of view).

\ssk {\bf Example.} {\em Grassmann geometries: the tangent geometry
of $(\Gras_E^F(W),\Gras_E^F(W))$ is simply
$(\Gras_{TE}^{TF}(TW),\Gras_{TE}^{TF}(TW))$, where $TW = W \oplus \eps W$, etc.,
with $\eps^2=0$ is constructed in the same way as the complexification of
a real vector space, replacing the condition $i^2=-1$ by $\eps^2=0$.
The action of $\PP \Gl_\K(W)$ is then replaced by the action of 
$\PP \Gl_{\K[\eps]}(TW)$. Even if there is no differentiable structure around,
everything behaves like a tangent object should do. 
Thus Grassmann geometries are indeed generalized projective geometries, and 
so are Lagrangian geometries.}

\msk \nin {\bf 3.4 Generalized polar geometries.} --
A {\em polarity} is an isomorphism of order $2$ of $(\XX^+,\XX^-)$ onto its
dual geometry $(\XX^-,\XX^+)$ and admitting at least one {\em non-isotropoic 
point} (see Section 1.3). 
A {\em generalized polar geometry over $\K$} 
is a generalized projective geometry over $\K$ together with a fixed polarity
$p:=p^+:\XX^+ \to \XX^-$, $p^-=p^{-1}$. Thus by definition
 the set $\MM:=\MM^{(p)}$ of
non-isotropic elements $x$ in $\XX^+$ is non-empty. 
The scalar $r=-1$ has the remarkable property that it is its own multiplicative
inverse. This property is the key for realizing on
the set $\MM^{(p)}$ the structure of a
{\em symmetric space}: one deduces from the
First Law that then the binary map
$$
\mu: \MM \times \MM \to \MM, \quad (x,y) \mapsto \mu(x,y):=\PPP_{-1}(x,p(x),y)=
(-1)_{x,p(x)}(y)
$$
is well-defined, and that is satisfies the following identities:
$\forall x,y,z \in \MM$,

\begin{description}
\item{(S1)} $\mu(x,x)=x$
\item{(S2)} $\mu(x,\mu(x,y))=y$
\item{(S3)} $\mu(x,\mu(y,z))=\mu(\mu(x,y),\mu(x,z))$
\end{description}

\nin
In other words, $\MM$ is equipped with a family of {\em symmetries}
$\sigma_x = \mu(x,\cdot)$ such that the $\sigma_x$ are
automorphisms  of order $2$ fixing $x$.  Following Loos
(\cite{Lo69}) we say that $\MM$
is a {\em symmetric space} if, moreover, the map $\mu$ is smooth and
the tangent map $T_x (\sigma_x)$
is the negative of the identity of the tangent space $T_x \MM$.
In fact, in our purely algebraic setting, this last property makes sense 
algebraically  because $\sigma_x$ is nothing but the dilation
$(-1)_{x,p(x)}$ which is just the negative of the identity map on the linear space
 $((p(x))^\top,x)$. By some abuse of language, we thus may call
 $\MM=\MM^{(p)}$ the {\em symmetric space of a generalized polar geometry}. 
(Under suitable assumtions, this is indeed a smooth symmetric space
in the differential geometric sense, see Theorem 3.6.1 below. 
If, moreover, the dimension is finite over $\K=\R$, then the topological
connected components are homogeneous symmetric spaces $G/H$, see \cite{Lo69}.)
Since the symmetric space clearly 
depends functorially on the generalized polar
geometry,  this correspondence
is called the {\em geometric Jordan-Lie functor} 
(see \cite{Be00}, \cite{Be02}).
We have already given in Section 1.3.3 some examples of symmetric spaces arising
in this way. 

\msk \nin {\bf 3.5 Null geometries.} --
An {\em absolute null  geometry over $\K$} 
is a generalized projective geometry over $\K$ together with a fixed
absolute null-system $n:\XX^+ \to \XX^-$ (that means that $n$ commutes with
all inner automorphisms of $(\XX^+,\XX^-)$).
Such geometries have several remarkable properties, some of which can be
used to give equivalent characterizations (cf.\ \cite{Be03}); for instance,
they admit ``inner polarities'':

\ssk {\bf Example.} {\em Fix two points $x,y \in \XX^+$ and
consider the {\em midpoint map} associating to every affinization
$\alpha \in \XX^+$ the geometric midpoint $\PPP_{\frac{1}{2}} (x,\alpha,y) \in
\XX^+$ of $x$ and $y$ in the affine part defined by $\alpha$.
Null geometries are characterized by the remarkable property that $x$ and
$y$ can be chosen such that every ``generic'' point can appear as
 geometric midpoint
of $x$ and $y$. Then the midpoint map extends to a bijection from
$\XX^-$ onto $\XX^+$, and  the Second Law now implies that this bijection is
a polarity, called an {\em inner polarity}. 
Just to see how surprising this property is, consider the case of
ordinary projective geometry: here the geometric midpoint of $x$ and $y$
always lies on the projective line spanned by $x$ and $y$, hence has a rather
special and non-generic position. 
However, if the projective space itself was already a projective line, then
indeed every point different from $x$ and $y$ can
play the role of a midpoint of $x$ and $y$. This corresponds to the fact, already
mentioned in the examples of Section 1.3.2, that among the usual projective spaces
only the projective line is a null geometry.}

\msk
\nin {\bf 3.6 The Jordan functor.} -- 
Recall that the {\em Lie functor} describes the
correspondence between Lie groups and Lie algebras. 
The Jordan analog of this correspondence
is described by the following result; in contrast to the Lie functor, the 
correspondence works equally well in arbitrary dimension and over 
general base fields and rings (we only have to assume that $2$ and $3$ are
invertible in $\K$). 

\bigskip
\nin {\bf 3.6.1 Theorem.} (The Jordan functor) {\em There are correspondences
(essentially bijections) between the following objects:

\begin{itemize}
\item[\rm (1)] 
connected  generalized projective geometries $(\XX^+,\XX^-)$ with base point
$(o^+,o^-)$, and Jordan pairs $ (V^+,V^-)$;
\item[\rm (2)] 
connected generalized polar geometries $(\XX^+,\XX^-,p)$ with base point
$o$, and Jordan triple systems $V$;
\item[\rm (3)]
 connected absolute null geometries $(\XX^+,\XX^-,n)$ with base point $o$, and
 Jordan algebras $V$ admitting a unit element $e$.
\end{itemize}
\nin
Moreover, these correspondences  are functorial in both directions.
(To be more precise, we have two functors $D$ ``differentiating at the base
point'' and $I$ ``integrating'' such that $D \circ I$ is the identity;
under certain restrictions such as finite dimensionality over a field, 
$D$ and $I$ are indeed equivalences of categories.)
 Moreover, under these correspondences,
\begin{itemize}
\item[\rm (i)]
the (algebraic) Jordan-Lie functor from Subsection 2.6 and
the geometric Jordan-Lie functor correspond to each other; 
\item[\rm (ii)]
inner ideals correspond to intrinsic subspaces containing the base point.
\end{itemize}
\nin
Finally, assume $\K=\R$, $\C$ or any other topological ring having a dense unit
group. Then $\XX^+$ and $\XX^-$ are smooth manifolds over $\K$
if (and only if) the Jordan pair satisfies a certain regularity condition
(called a {\em continuous quasi inverse Jordan pair} in \cite{BeNe05}). If we add
a continuous polarity $p$ to these data, then the corresponding symmetric space
$\MM^{(p)}$
is an open submanifold of $\XX^+$, and its symmetric space structure
is smooth.
}

\bigskip {\bf Examples.} {\em Under the correspondence from the theorem,
\begin{itemize}
\item
 Grassmann geometries $(\Gras_E^F(W),\Gras_F^E(W))$ correspond to
the Jordan pair  of ``rectangular matrices'' $(\Hom(E,F),\Hom(F,E))$.
If $E \cong F$, then this is a null geometry corresponding to the Jordan algebra
$\End(E)$.
\item
Lagrangian geometries of a  symplectic form are null geometries and
 correspond to Jordan algebras of symmetric matrices.
\item
Lagrangian geometries of a symmetric neutral form correspond to 
Jordan pairs of skew-symmetric matrices. They are null-geometries if these
matrices are of even size ($2n \times 2n$). For $\K=\C$,
Lagrangian geometries of a Hermitian neutral form correspond
to Jordan pairs of skew-Hermitian matrices; they
are always
null geometries, corresponding to the isomorphism between skew-Hermitian and
Hermitian matrices and to the Jordan algebra structure on the latter.
\item
Projective quadrics carry a structure of a null geometry that corresponds to
the spin-factor (see Item (4) in the example of 2.1).
\end{itemize}
}

\ssk
\nin {\bf Remarks about the theorem and its proof.} 
As for the correspondence between Lie groups and Lie
algebras, one has to give two constructions that are inverse to each other:
starting with a geometry, we get the associated infinitesimal object by
``differentiation''. For Lie groups, differentiating one, 
one simply gets the tangent space without any
useful information (the Lie algebra $\g$ with addition, but no bracket);
differentiating twice, one gets the Lie bracket, and finally one needs a third
order argument to prove the Jacobi identity. In the present case
the situation is quite similar (see \cite{Be02}).

\ssk
The inverse construction is fairly easy:
roughly, starting from a Jordan pair $(V^+,V^-)$, one uses the
Kantor-Koecher-Tits construction described in Section 2.2 to define the associated 
$3$-graded Lie algebra $\g$. Since elements of $\ad(\g_1)$ and $\ad(\g_{-1})$ are
$2$-step nilpotent, their exponential is just a quadratic polynomial, so that
we have well-defined groups $U^\pm =\exp(\ad(\g_{\pm 1}))$.
The subgroup $G = \langle U^+,U^- \rangle$ of $\Aut(\g)$ generated by these two
groups is called the {\em elementary projective group of the Jordan pair}.
Let $H \subset G$ be the stabilizer of the $3$-grading (``diagonal matrices in $G$'')
and $P^\pm := H U^\pm$ (``parabolics''; this is a semidirect product).
Then $(\XX^+,\XX^-)=(G/P^-,G/P^+)$, with transversality defined by the 
$G$-orbit of the canonical base point $(eP^-,eP^+)$, is the geometry sought for.
The hard part is now to verify that
(PG1) and (PG2) indeed hold and that the construction is functorial (see 
\cite{Be02}).

\ssk
Part (3) on null geometries is proved in \cite{Be03}, and statement (ii) on
intrinsic subspaces in \cite{BL06}. 
The final statement on smooth structures has been obtained in joint work with
K.-H.\ Neeb (\cite{BeNe04}, \cite{BeNe05}); see also \cite{Be07}.
The setting of differential calculus,  smooth manifolds and Lie groups
 over topological base fields and rings has been
developed in \cite{BGN03} -- see also \cite{Be06}, Chapter 1, for the basic facts;
we hope to convince the reader that the
 resulting theory really is much simpler than 
``usual'' differential calculus in Banach or Fr\'echet spaces. 
It
 covers, in particular, the interesting cases $\K=\Q_p$ (the $p$-adic
numbers), $\K=\R \times \R$ (para-complex numbers) and $\K=\R[\eps]$
(dual numbers). 
An approach to differential geometry, Lie groups and
symmetric spaces in this general framework is worked out in \cite{Be06};
 much of the algebraic framework developed there
has been designed aiming at applications in situations like the ones
discussed here. 
 
\ssk {\bf Example.} {\em
A {\em Banach Jordan pair} is a pair of Banach spaces with a continuous
trilinear Jordan pair structure. In this case the conditions from \cite{BeNe05}
are always fulfilled, so that the corresponding geometry
$(\XX^+,\XX^-)$ is indeed a pair of smooth
(Banach-)manifolds. (This can also be proved by more conventional functional
analytic methods, see the monograph \cite{Up85}.) For instance,
full or Hermitian algebras of continuous operators are Banach Jordan pairs,
and hence the geometry associated to $\Herm(\HH)$ is a smooth manifold,
in fact isomorphic to $\UU(\HH)$.
There is a huge literature on Banach Jordan structures, see references in
\cite{AlSch01}, \cite{Io03} or \cite{Up87}.
}

\section{Comments and prospects}

As said in the introduction, the theory exposed in this paper is purely
mathematical; the description by using terms borrowed from the language
of physics may be seen as a game without any relation to the ``real
world''. Be this as it may, I would like to put forward some arguments why I think
that it is interesting to play this game and maybe  to pursue it
even further: 

\msk {\bf (1)}
First of all, it is certainly not the only, but at least
one possible interpretation of the ``Jordan Ansatz'',
and hence it does cover the standard setting of  quantum mechanics
(the Jordan algebra $\Herm(\HH)$):
thus it should  have {\it some} meaning, be it relevant or not.
%

\msk {\bf (2)}
Apart from the original motivation by the Jordan Ansatz,
our setting incorporates other viewpoints that have shown up in the
search for foundations of quantum mechanics:
\begin{itemize}
\item
the fundamental r\^ole of projective geometries (\cite{Va85}, p.\ 6: 
``... quantum mechanical systems are those whose logics form some sort
of projective geometries'', \cite{BH01}:
``...any specific feature of projective geometry gives rise to a
physically realisable characteristic of quantum mechanics''); 
\item
linearity (\cite{AS98}:
``Perhaps the habitual linear structures of quantum mechanics are analoguous
to the inertial rest frames in special relativity...'');
\item
duality
(\cite{Om94}, p.\ 527: 
``The pure philosopher may start from a postulated unity and call it
Being. He may then concede the necessity of distinguishing two modes of
being and call them reality and logos or whatever 
else...  A physicist groping with his science is after all
following the same path...'');
\item
quantum non-locality (see the comparison with twistor theory
in Section 1.4.2);
%
\item
Hermitian symmetric spaces (see, e.g., the paper
 ``The pure state space of quantum mechanics as Hermitian symmetric
space''  \cite{CGM03}.
One should not underestimate the fact that modern Jordan theory
is intimately related to the theory of finite and infinite dimensional
Hermitian symmetric spaces, by work of Koecher, Loos, Kaup, Upmeier and others;
cf.\ \cite{Ka83}, \cite{Up85} and
the ``Colloquial Survey of Jordan Theory'' in \cite{McC04}. 
This aspect has not been noticed at all in 
\cite{CGM03}.)
\item
the search for using ``exceptional'' groups and geometries in physics
(Jordan theory is deeply mixed into the structure of Freudenthal's
``magic square'' which is a main source of exceptional
geometries; cf.\ also references in \cite{Io03}.)
\end{itemize}
\nin If one strives for unity of mathematics,  it is very satisfying
to realize that all these aspects can be incorporated, without mutilating
any of them, in a common framework.

\msk{\bf (3)}
I consider ``non-associative geometry'' as some sort of natural counterpart
of ``non-commutative geometry'' (see Subsection 3.1.3 and \cite{Be07};
as usual ``non'' means here: ``not necessarily'').\footnote{As I learned
later, the term
``non-associative geometry'' has previously been introduced by L.V.\
Sabinin \cite{NS00} in the more specific context of {\em quasigroups}
and {\em loops}. Symmetric spaces are prominent examples of such 
structures, but as far as I know, Jordan algebraic structures cannot
be interpreted in this context.}
As I tried to explain in my discussion of the quotation from \cite{AlSch01},
even if at the end we will be forced to  return to some ``associative
geometry''  for
describing quantum mechanics, the decomposition of the associative product
in its Jordan and Lie part somehow seems to correspond to the fundamental
problem of the coexistence of the $\bf U$-evolution and the state reduction $\bf R$.
Therefore it might be useful to widen the scope from associative to 
non-associative structures.

\msk{\bf (4)}
Be our universe unique or not, a mathematician  has the
natural desire to look at it as belonging to a {\it category},
so that usual categorial notions should apply to it. 
This means that he would like to understand the universe by
 its {\it properties} and not by a construction such as
``take an infinite dimensional separable Hilbert space 
and do this and that...''. 
We encounter the same problem already on the level of ordinary
projective spaces: one can define them by the usual
 construction (``take the rays
in a vector space...''), or by intrinsic properties -- when doing 
the latter,
all modern authors more or less
follow the famous model of Hilbert's ``Grundlagen der Geometrie''
where the {\it incidence axioms} of projective geometry were shown to
be the intrinsic geometric properties sought for.
We simply propose to replace here ``incidence axioms'' by other 
foundational properties, namely by ``laws'', in the sense of Sections 1.2.2 and 3.2.
It turns out that, unlike incidence structures,
algebraic laws are very flexible and can be
adapted to a great variety of situations.
For instance, notions of {\it direct products} and {\it bundles} (see below)
 exist in our
setting, whereas they are in general not compatible with interesting incidence 
axioms (the direct product of two ordinary
projective spaces is no longer an ordinary projective space!).
In some sense, it seems to me that this is the deeper mathematical reason why
the incidence geometric approach to quantum mechanics,
due to Birkhoff and von Neumann,  
 has been gradually abandoned, in spite of
its great mathematical beauty (see the book \cite{Va85}).

\msk{\bf (5)} 
I already mentioned (Section 1.5) the problem that
``tensor products of Jordan algebras'' do not exist and hence it is not
clear at all how many-particle systems should be modelized in
our geometric approach. However, thanks to the flexibility of algebraic
laws just mentioned, the situation is not as hopeless as
it might look. 
For several reasons, it seems to me that the suitable geometric setting 
replacing the tensor product construction from usual quantum mechanics
should be related to a notion of {\em vector bundles in the category of generalized
projective geometries}. In fact, ``vector bundles in the category of symmetric
spaces'' have been introduced and studied in \cite{Did06} and \cite{BeD07}:
essentially, a vector bundle in some geometric category is 
a vector bundle $F$ over a base $M$ such that both $F$ and $M$ are objects
of the category and such that some natural compatibility
conditions are satisfied. 
On the infinitesimal level, the corresponding notion is the one of
 {\em general representation} or {\em $\m$-module}
for the tangent object $\m$ of the base $M$.
For instance, in the category of Lie groups we get the usual notion
of representation or $G$-module of the base $G$.
Now, similarly as in the category of Lie groups, for symmetric spaces
and generalized projective geometries, 
there exist notions replacing tensor products for such representations.
As a variant of this, one may also leave the category of vector bundles
and modelize many-particle systems by {\it multilinear bundles} which
have been introduced in \cite{Be06} precisely with the aim to 
replace tensor products of vector bundles by more geometric notions.
In any case, it is tempting, by employing  the language of
$\m$-representations, to associate ``atoms'' with simple geometries or
irreducible representations, ``molecules'' with suitable extensions of
simple geometries, and, on the opposite end, to interprete ``classical 
mechanics'' via function geometries, that is, certain continuous direct products
of the standard fiber which has
not much interesting internal structure.

\msk{\bf (6)}
 The fact that we may think of generalized projective
geometries both classically and quantum-mecanically is rather puzzeling
(see example in Subsection 1.2.3): 
the conformal compactification of Minkowski-space as well as the
geometry of quantum mechanics  are examples of generalized projective
geometries. 
Is this an accident? Some speculations about this question
can be found in \cite{Cas77}.\footnote{\footnotesize
For a popular account see p.\ 404 in: 
von Weizs\"acker, C.\ F., {\em Aufbau der Physik}, Hanser-Verlag,
M\"unchen 1985.}
%
%

\msk {\bf (7)}
The mathematical similarity between compactified Minkowski space and
the generalized projective geometry of quantum mechanics suggests that
it may be necessary to
carry these ideas even further and to go beyond the category of generalized
projective geometries: 
namely, comparing with the
historical development of theories
 from Newtonian mechanics via Special Relativity to 
General Relativity, Newtonian mechanics would correspond to the standard
Hilbert space formulation of quantum theory, Special Relativity (in compactified
Minkowski space)
would correspond to our hypothetical ``generalized projective'' formulation, 
and hence General Relativity should correspond to
an even more hypothetical generalization in terms of
 geometries
that are ``modelled on generalized projective geometries'', but are no
longer ``conformally flat'' (in a suitable sense; recall that compactified
Minkowski space is conformally flat). 
In finite dimension,  geometries modelled on certain homogeneous spaces $G/P$
 have been studied by Elie Cartan using what is nowadays called a 
{\em Cartan connection}
(which generalizes the projective and conformal connections;
see \cite{Sh97} for a modern  presentation).
As far as I know, Jordan theory or more general non-associative
algebra have not yet been systematically used
as an approach to study the corresponding Cartan geometries, but certainly
this would open the way for a theory of their infinite-dimensional
generalization. 
Summing up, if the  analogies mentioned above have some meaning, 
it could be hoped that Jordan geometry gives some hints on what the last
two items in the following matrix might be:
$$
\begin{matrix}
\mbox{\em geometry:} &
\mbox{linear; affine} &  \mbox{projective} &  \mbox{manifold}  \cr
\mbox{\em mechanics:} &
\mbox{classical} &  \mbox{special relativistic} & \mbox{general 
relativistic}  \cr
\mbox{\em quantum theory:} &
\mbox{Hilbert space q.m.} &  \mbox{projective q.m. ?} &  
\mbox{Cartan geometric q.m. ??}  \cr
\end{matrix}
$$
\bigskip
$$
\phantom{xx}
$$



%

%

}

\end{document}